\documentclass[12pt]{article}
\usepackage[breaklinks=true,colorlinks=true,backref,pagebackref]{hyperref}

\usepackage{epsfig}
\usepackage{amsmath,amssymb,amsfonts,graphicx}
\usepackage{youngtab}
\usepackage[mathscr]{eucal}
\usepackage{multicol}
\usepackage{color}
\usepackage{pst-coil}
\usepackage{colortbl}
\usepackage[all]{xy}
\usepackage{youngtab}

\def\a{\alpha}
\def\b{\beta}
\def\d{\delta}
\def\g{\gamma}

\def\eps{\epsilon}
\def\ve{\varepsilon}
\def\vt{\vartheta}

\def\stareq{\stackrel{*}{=}}

\def\rstar {{\color{red}(*)}}

\begin{document}

\title{The Kummer tensor density in
  electrodynamics and in gravity}

\author{Peter Baekler$^{1}$, Alberto Favaro$^{2}$, Yakov Itin$^{3}$,\\
  Friedrich W. Hehl$^{4,*}$\\
  $^{1}$Univ.\ of Appl.\ Sciences, 40474 D\"usseldorf, Germany\\
  $^{2}$Inst.\ Physics, Carl-von-Ossietzky-Univ.,\\
  26111 Oldenburg, Germany\\
  $^{3}$ Inst.\ Mathematics, Hebrew Univ. of Jerusalem {\it and}\\
  Jerusalem College of Technology, Israel\\
  $^{4}$Inst.\ Theor.\ Physics, Univ. Cologne,\\ 50923 K\"oln,
  Germany, {\it and} Dept.\ Physics \& Astron.,\\ Univ.\ of Missouri,
  Columbia, MO 65211, USA\\ $^{*}$ Corresponding author: {\it email
    hehl@thp.uni-koeln.de}}

\date{28 Sept 2014, {\it file KummerTensor33.tex}}

\maketitle

\begin{abstract}
  Guided by results in the premetric electrodynamics of local and
  linear media, we introduce on 4-dimensional spacetime the new
  abstract notion of a Kummer tensor density of rank four, ${\cal
    K}^{ijkl}$. This tensor density is, by definition, a cubic
  algebraic functional of a tensor density of rank four ${\cal
    T}^{ijkl}$, which is antisymmetric in its first two and its last
  two indices: ${\cal T}^{ijkl}=-{\cal T}^{jikl}=-{\cal
    T}^{ijlk}$. Thus, ${\cal K}\sim {\cal T}^3$, see
  Eq.(\ref{KummerDef}). (i) If $\cal T$ is identified with the
  electromagnetic response tensor of local and linear media, the
  Kummer tensor density encompasses the generalized {\it Fresnel wave
    surfaces} for propagating light. In the reversible case, the wave
  surfaces turn out to be {\it Kummer surfaces} as defined in
  algebraic geometry (Bateman 1910). (ii) If $\cal T$ is identified
  with the {\it curvature} tensor $R^{ijkl}$ of a Riemann--Cartan
  spacetime, then ${\cal K}\sim R^3$ and, in the special case of
  general relativity, ${\cal K}$ reduces to the Kummer tensor of Zund
  (1969). This $\cal K$ is related to the {\it principal null
    directions} of the curvature. We discuss the properties of the
  general Kummer tensor density. In particular, we decompose $\cal K$
  irreducibly under the 4-dimensional linear group $GL(4,R)$ and,
  subsequently, under the Lorentz group $SO(1,3)$. 
%
\end{abstract}
\tableofcontents
\newpage
\section{Introduction}

\subsection{Fresnel surface}

We consider electromagnetic waves propagating in a homogeneous,
transparent, dispersionless, and nonconducting crystal. The response
of the crystal to electric and magnetic perturbations is assumed to be
{\it local} and {\it linear.} The permittivity tensor
$\varepsilon^{ab}$ ($a,b=1,2,3$) of the crystal\footnote{The position
  of the indices are chosen always in accordance with the conventions
  of premetric electrodynamics, see \cite{Post,Birkbook,Hehl:2007ut}.}
is anisotropic in general, and the same is true for its {\it
  im\/}permeability tensor $\mu^{-1}_{ab}$. Such materials are called
special bi-anisotropic (see
\cite{Serdyukov:2001,Ismobook,Sihvola-meta}). If $\varepsilon^{ab}$
and $\mu^{-1}_{ab}$ are assumed to be symmetric, these bi-anisotropic
materials are characterized by 12 independent parameters. In many
applications, however, $\mu^{-1}_{ab}$ can be considered approximately
to be isotropic $\mu^{-1}_{ab}=\mu^{-1}_0g_{ab}$; here $g_{ab}$ is the
3-dimensional Euclidean metric tensor. Such cases were already studied
experimentally and theoretically in the early 19th century, before
Maxwell recognized the electromagnetic nature of light in 1862 (see
\cite{DarrigolOptics}). To carry out these investigations on
geometric optics, one used the notions of light ray and of wave
vector, and one was aware (Young, 1801)\footnote{Thomas Young
  (1773--1829), English mathematician and physicist.} that light was
transverse and equipped with a polarization vector.

At each point inside a crystal, we have a ray vector and a wave
covector (one-form). It is then possible to determine the {\it ray
  surface} and its dual, the {\it wave surface,} for visualizing how a
pulse of light is propagating. The ray surface was first constructed
by Fresnel (1822) and is conventionally called {\it Fresnel surface,}
an expression also used for the wave surface, see the popular
introduction by Kn\"orrer \cite{Knoerrer}. Since the symmetric
permittivity tensor can be diagonalized, the Fresnel surface is
described by 3 principal values. In the case when all of them are
equal, the Fresnel surface is an ordinary 2-dimensional sphere. For
two unequal parameters, the surface is the union of two shells, a
sphere and an ellipsoid. These two shells touch at two points. In
Fig.1, we display such a surface for a crystal with {\it three
  different principal} values of the permittivity:
  $(\ve^{ab})=\left(\begin{smallmatrix}\ve^{1}&0&0\\0&\ve^{2}&0\\
      0&0&\ve^{3}\end{smallmatrix}\right)$ and $(\mu^{-1}_{ab})=
  \mu^{-1}_0\left(\begin{smallmatrix}1&0&0\\0&1&0\\
      0&0&1\end{smallmatrix}\right)$. It is  \newpage
\includegraphics[width=0.05cm]
{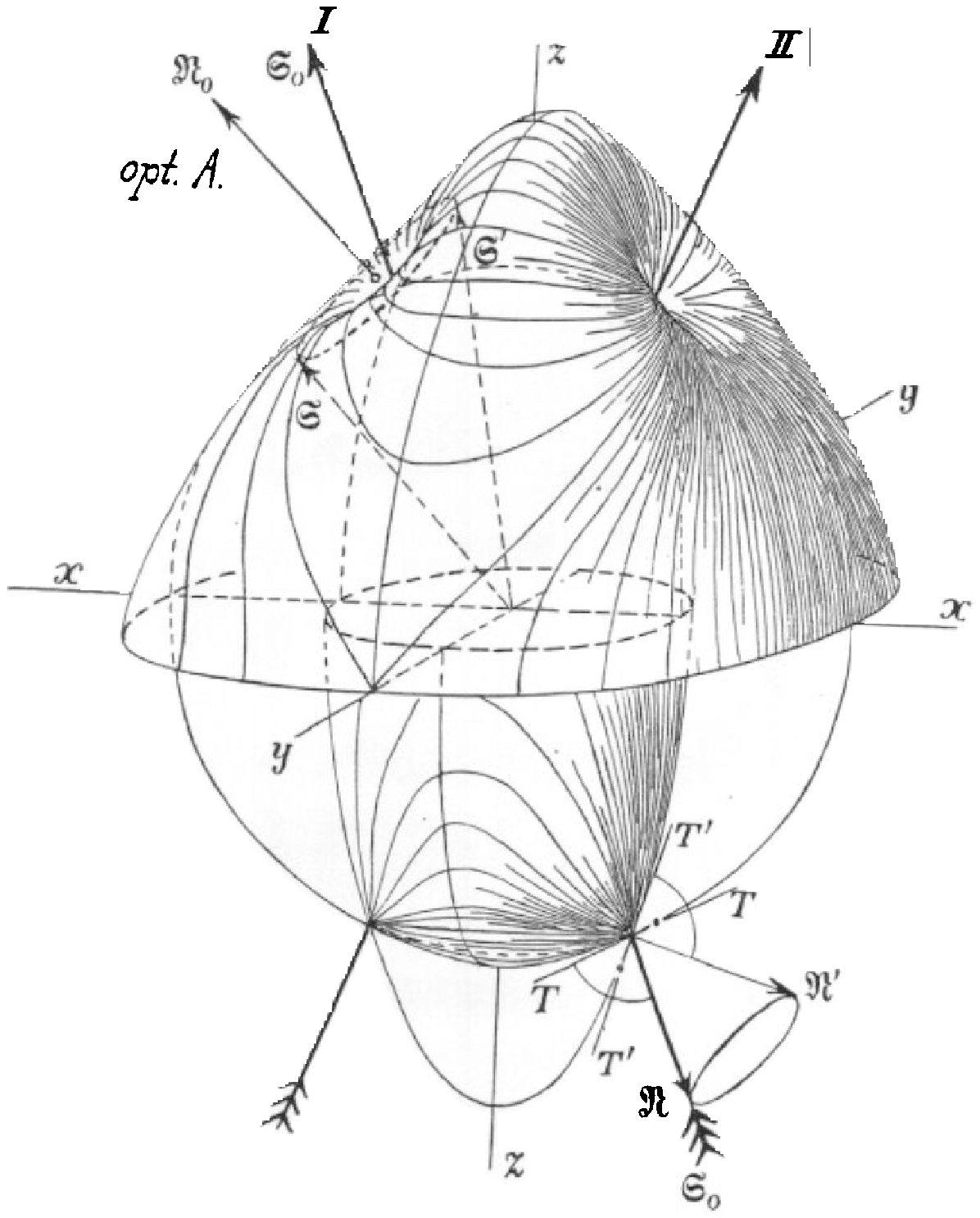}\bigskip
\vspace{300pt}

\begin{footnotesize}
Figure 1. Fresnel wave surface as the specific {\it quartic} surface
\begin{eqnarray}\label{FresnelQuadric}
&&(\a^2x^2+\b^2y^2+\g^2z^2)(x^2+y^2+z^2) \nonumber\\
&&\hspace{10pt}-\left[\a^2(\b^2+\g^2)x^2+\b^2(\g^2+\a^2)y^2+\g^2(\a^2+\b^2)z^2
 \right]+\a^2\b^2\g^2=0\,,\nonumber
\end{eqnarray}
with the 3 parameters
$\a:=c/\sqrt{\ve_{1}},\;\b:=c/\sqrt{\ve_{2}},\;\g:=c/\sqrt{\ve_{3}}$ and
$c=$ vacuum speed of light (from Schaefer \cite{Schaefer}, image
by J.~Jaumann).  The upper half depicts the exterior shell
with the funnel shaped singularities, the lower half the inner
shell. The two optical axes are denoted by ${\bf I}$ and $\bf II$. The
two shells cross each other at four points forming cusps. The wave
vectors are denoted by $\frak{N}$ and $\frak{N}'$. All essential facts
of crystal optics are encoded in this drawing. Such quartic Fresnel
surfaces and their non-trivial generalizations are at center stage of
our article.
\end{footnotesize}

a union of two shells that meet at 4 singular points. As already
recognized by Hadamard \cite{Hadamard}, the wave surfaces are the
characteristics of the corresponding partial differential equations
describing the wave propagation.

The classical 3-parameter Fresnel surface is naturally generalized
when the anisotropy of the impermeability $\mu^{-1}_{ab}$ is taken
into account. Removing also the diagonalizability requirement, one
deals with {\it generalized Fresnel surfaces} of less than 12
parameters. Such surfaces were derived in terms of positive definite
symmetric dyadics by Lindell \cite{Lindell1973}, see also
\cite{KiehnFresnel}.

Moreover, it seems to be rather natural to extend $\varepsilon$ and
$\mu$ to asymmetric tensors, which emerge if dissipative processes are
involved.  Recently, one of us \cite{Itin:2010} derived a tensorial
expression of such a generalized 18-parameter Fresnel surface. The
derivation does not require the corresponding matrices to be real,
symmetric, positive definite, or even invertible.

We here will derive such tensorial expressions for generalized Fresnel
surfaces by proceeding differently. We include first magnetoelectric
effects and subsequently look for the corresponding 4-dimensional
relativistic covariant generalizations of the 3-dimensional
permittivity and impermeability tensors.

\subsection{Magnetoelectricity}

In the 1960s, substances were found that, if exposed to a magnetic
field $B^a$, were electrically polarized $D^a=\a^a{}_bB^b$ and,
reciprocally, if exposed to an electric field $E_a$, were magnetized,
$H_a=\b_a{}^bE_b$, see O'Dell \cite{O'Dell}. These are small effects of the
order $10^{-3}\sqrt{\varepsilon_0/\mu_0}$, or smaller.

Such materials are characterized by the constitutive moduli $
\ve^{ab}$, $\mu^{-1}_{ab}$, $\a^a{}_b$, $\b_b{}^a$, which can be
accommodated in a $6\times 6$ matrix. Originally, all these moduli
were assumed to obey the {\it symmetry} conditions
\begin{equation}\label{symmetric} \ve^{ab}= \ve^{ba}\,,\qquad
  \mu^{-1}_{ab}=\mu^{-1}_{ba}\,,\qquad\text{and}\qquad\a^a{}_b=-\b_b{}^a\,,
\end{equation} yielding
altogether $2\times 6+9=21$ moduli characterizing a material.

Seemingly, Bateman in 1910 \cite{Bateman_1910} was the first to
investigate such materials. He studied electromagnetic wave
propagation in Maxwell's theory. His constitutive relations, in his
notation, were
\begin{eqnarray}\label{constBateman}
  -B_1&=&\kappa_{11}H_1+\kappa_{12}H_2+\kappa_{13}H_3+\kappa_{14}D_1
  +\kappa_{15}D_2+\kappa_{16}D_3\,,\nonumber\\
  &&\hspace{-5pt} ...\hspace{28pt}...\hspace{35pt}...\hspace{35pt}...
  \hspace{35pt}...\hspace{35pt}...\nonumber\\
  E_3&=&\kappa_{61}H_1+\kappa_{62}H_2+\kappa_{63}H_3+\kappa_{64}D_1
  +\kappa_{65}D_2+\kappa_{66}D_3\,.
\end{eqnarray}
Bateman assumed the existence of 21 moduli $\kappa_{IJ}=\kappa_{JI}$,
for $I,J=1,...,6$, believing that {\it ``These conditions} [the
constitutive relations] {\it may not correspond to anything occurring
  in nature; nevertheless their investigation was thought to be of
  some mathematical interest on account of the connection which is
  established between line-geometry and the theory of partial
  differential equations.}'' Then, {\it ``The general Kummer's surface
  appears to be the wave surface for a medium of a purely ideal
  character...''}

Bateman's 21 independent moduli $\kappa_{IJ}$---in contrast to what he
thought---are not only a mathematical abstraction. Rather, they do
have applications in physics, since they coincide with the 21 moduli
displayed in (\ref{symmetric}).

Of course, nowadays it is desirable to display the constitutive law
 (\ref{symmetric}) or (\ref{constBateman}), respectively, in a
4-dimensional covariant and premetric way, see
\cite{Post,O'Dell,Birkbook}; Whittaker \cite{Whittaker}, Vol.~2,
pp.~192--196, provides a short history of the premetric program. We
collect the fields $D^a$ and $H_a$ in the 4d electromagnetic
excitation tensor density ${\cal H}^{ij}\;(=\!\!-{\cal H}^{ji})$ and
the fields $E_a$ and $B^a$ in the electromagnetic fields strength
tensor $F_{ij}\;(=\!\!-F_{ji})$, with the coordinate indices
$i,j,k,...=0,1,2,3$. Then, Maxwell's equations, with the electric
current density ${\cal J}^j$, read
\begin{eqnarray}\label{Maxwellcomp}
  \partial_j{\cal H}^{ij}={\cal J}^j\,,\qquad  \partial_{[i} F_{jk]} = 0\,. 
\end{eqnarray}
We assume that the crystal considered above responds locally and
linearly if exposed to electromagnetic fields. As
a consequence, the constitutive law, with the {\it electromagnetic response
tensor density} $\chi^{ijkl}(x)$, reads as follows:
\begin{eqnarray}\label{const} {\cal H}^{ij}=\frac 12 \chi^{ijkl}F_{kl}\,,
  \qquad\text{where}\qquad\chi^{ijkl}=- \chi^{jikl}\,,\quad \chi^{ijkl}=-
  \chi^{ijlk}\,.
\end{eqnarray}
The response tensor density $\chi$ can be mapped to a $6\times 6$
matrix with 36 independent components. This matrix can be decomposed
in its trace-free symmetric part (20 independent components), its
antisymmetric part (15 components), and its trace (1 component). On
the level of $\chi$, we have then the irreducible decomposition under
$GL(4,R)$:
\begin{eqnarray}\label{chidec}
  \chi^{ijkl}&=&\,^{(1)}\chi^{ijkl}+
  \,^{(2)}\chi^{ijkl}+
  \,^{(3)}\chi^{ijkl}\,,\\ \nonumber 36
  &=&\hspace{11pt} 20\hspace{15pt}\oplus \hspace{11pt}15\hspace{11pt}
  \oplus \hspace{25pt}1\,,
\end{eqnarray}
see \cite{Birkbook,Ismobook,Hehl:2007ut}. The third part, the {\it
  axion} part, is totally antisymmetric and as such proportional to
the Levi-Civita symbol, $ ^{(3)}\chi^{ijkl}:= \chi^{[ijkl]}
={\a}\,{\epsilon}^{ijkl}$ (see also
\cite{Sihvola,RaabSihvola,Lindell-Sihvola}). The second part, the {\it skewon}
part, is defined according to $ ^{(2)}\chi^{ijkl}:=\frac
12(\chi^{ijkl}- \chi^{klij})$, see also \cite{Obukhov:2004,Itin:2013}.  If the
constitutive equation can be derived from a Lagrangian, which is the
case as long as only reversible processes are considered, then
$^{(2)}\chi^{ijkl}=0$. This corresponds to Bateman's symmetry
postulate $\kappa_{[IJ]}=0$. The {\it principal} part
$^{(1)}\chi^{ijkl}$ fulfills the symmetries $ ^{(1)}\chi^{ijkl}=
{}^{(1)}\chi^{klij}$ and $^{(1)}\chi^{[ijkl]}=0$.

\subsection{Wave propagation: Tamm-Rubilar tensor density $ {\cal
    G}^{ijkl}[\chi]$}

In Sec.1.1, we studied the Fresnel wave surfaces of crystals with an
anisotropic dielectric constant $\ve^{ab}$. Let us generalize these
considerations to the response tensor density $\chi^{ijkl}$ of
(\ref{const}), with its 36 independent components. The 4d wave covector
is denoted by $q_i=(\omega,-{\mathbf q})$, with the frequency $\omega$ and the
3d wave vector $\mathbf q$. As known from the literature, see
\cite{Birkbook} and the references given there, one arrives at the
generalized Fresnel equation, which is again {\it quartic} in the wave
covector:
\begin{equation}\label{Fresnel} {\cal
    G}^{ijkl}[\chi]\,q_iq_jq_kq_l=0\,.
\end{equation}
The {Tamm-Rubilar (TR) tensor density} $\cal G$ is totally symmetric,
${\cal G}^{ijkl}[\chi]={\cal G}^{(ijkl)}[\chi]$, and has 35 independent
components.\footnote{Since by (\ref{Fresnel}) the overall factor in the
  definition of $\cal G$ is conventional, $\cal G$ effectively has
  only 34 independent components.} The explicit definition reads, see
\cite{Birkbook}, Eq.(D.2.22),
\begin{equation}\label{G4}   
{{\cal G}^{ijkl}[\chi]:=\frac{1}{4!}\,{\epsilon}_{mnpq}\, 
  {\epsilon}_{rstu}\, {\chi}^{mnr(i}\, {\chi}^{j|ps|k}\, 
  {\chi}^{l)qtu }\,,} 
\end{equation} 
that is, it is {\it cubic} in $\chi$. The totally antisymmetric
Levi-Civita tensor density is denoted by $\epsilon_{ijkl}=\pm1,0$, and
the parentheses $\,^{(\;)}$ mark symmetrization over the enclosed
indices; however, the indices standing between the two vertical
strokes $^{|\hspace{5pt}|}$, here $p$ and $s$, are excluded from the
symmetrization process, see \cite{Schouten_1954}. It turns out that
$\cal G[\chi]$ depends only on $^{(1)}\chi$ and $^{(2)}\chi$; the
axion piece $^{(3)}\chi$ drops out.

Itin \cite{Itin:2009aa} developed a new and generally covariant method
for deriving the generalized Fresnel equation (``dispersion
relation''). His explicit expression for the TR-tensor density looks
different from (\ref{G4}). However, Obukhov
\cite{Obukhov/Itin}, by straightforward but cumbersome algebra, was
able to show that the results are {\it equivalent.} In this context,
Obukhov was able, by using the double dual, to put (\ref{G4}) into
what Itin called ``Probably the most symmetric form...'' of the
TR-tensor density, namely\footnote{
By using the duality operators, Eq.(\ref{G4}) reads
$\,{\cal G}^{ijkl}[\chi]=\frac{1}{3!}\,^\diamond {\chi}_{ab}{}^{c(i}\, 
{\chi}^{j|ad|k}\,  {\chi}^\diamond\,^{l)b}{}_{cd}$ .
Schuller et al.\ \cite{Schuller:2009hn}, in their work on the area
metric, took also our original version of the TR-tensor, as displayed
in (\ref{G4}).}
\begin{equation}\label{G4doublediamond}   
{\cal G}^{ijkl}[\chi]=\frac{1}{3!}\, \chi^{a(i j|b}\,
{^\diamond}\chi^\diamond{}_{\hspace{-2pt}acbd}\,\chi^{c|kl)d}\,;
\end{equation} 
here we have the left-dual
$^\diamond\chi_{ij}{}^{kl}:={\scriptstyle{\frac
    12}}\eps_{ijmn}\chi^{mnkl}$, the right-dual
$\chi^{\diamond\,ij}{}_{kl} :={\scriptstyle{\frac
    12}}\chi^{ijmn}\eps_{klmn} $, and, thus, the double dual
${^\diamond}\chi^\diamond{}_{\hspace{-2pt}ijkl}=\frac 14
\eps_{ijmn}\chi^{mnrs}\eps_{klrs}$. This version was derived directly
by Lindell \cite{Lindell:2005} by using a dyadic calculus, see the
discussion of Favaro \cite{Favaro:2012}.

The subject of our article is the Kummer tensor density ${\cal
  K}^{ijkl}[\chi]$ that we find from (\ref{G4doublediamond}) by
dropping the factor $1/3!$ and the symmetrization parentheses
$^{(..|..|..)}$.  It should be well understood that the expressions of
the TR-tensor density in equations (\ref{G4}) and
(\ref{G4doublediamond}) are {\it equivalent} to each other. However,
if we generalize the TR-tensor density in Sec.2, it is decisive that
we start exactly from equation (\ref{G4doublediamond}).

\subsection{Kummer surface and optics}

We already saw that, as one examines increasingly structured
materials, the Fresnel images of the wave surfaces become more
complicated, and depend on a wider set of characteristic
parameters. However, equation \eqref{Fresnel} specifies that, even in
the general case, we are dealing with a \textit{quartic} hypersurface
in the 4d space of wave covectors $q_i=(\omega,-\mathbf{q})$. Plots
such as Figure 1 are derived by taking the section $\omega=1$, and
represent the inverse phase velocity of light.

In 1864, Kummer \cite{Kummer:1864,Kummer:1864a} discovered a family of
quartic surfaces that would later play a major role in the development
of algebraic geometry. Kummer surfaces belong to the 3d real
projective space ${R}P^3$, but can be derived from hypersurfaces in
the 4d space ${R}^4-\{0\}$, by taking a section. This is usually
regarded as a change of coordinates, from the homogeneous
$(t,\mathbf{x})$ to the inhomogeneous $(1,\mathbf{x}/t)$, see also the
explanations and visualizations in Lord \cite{LordProjGeom}.

As we establish below, for a dispersionless linear medium that is only
required to have zero skewon part, the Fresnel surface coincides with
the general Kummer surface. Is this equality a big surprise? Perhaps,
not. Kummer's original motivation for investigating quartic surfaces
came from optics. He wanted to improve Hamilton's geometric optics
(1832) and was interested, for example, in how the atmosphere of the
Earth modifies the image of the Sun or a planet \cite{Kummer:1863}. In
fact, he started to examine ray bundles of the second order, that is,
ray bundles that have at one point two independent rays, like in a
birefringent crystal. Given that rays in 4d space are projected in a
3d space, the tool for this analysis was projective
geometry.

The surfaces considered in the articles of 1864
\cite{Kummer:1864,Kummer:1864a} were named after Kummer by Hudson, who
wrote an authoritative book \cite{Hudson_1903} on the
subject. However, let us emphasize that already, in 1846, Cayley
\cite{Cayley} found special Kummer surfaces---so-called
tetrahedroids---whilst investigating light propagation, see also
\cite{Hudson_1903}.

Let us turn to the visualization of Kummer surfaces.  In Hudson
\cite{Hudson_1903} one can find, in suitably chosen coordinates, a
parametrization of the general Kummer surface. However, we will take a
parametrization of the Mathematica library \cite{Mathematica}: the
Kummer surfaces are a family of {quartic} surfaces. If we specify 4
coordinates $(w,x,y,z)$, a representation is given by
\begin{eqnarray}
(x^2+y^2+z^2-\mu^2w^2)^2-\lambda pqrs=0, \qquad\text{where}\qquad
\lambda:=\frac{3\mu^2-1}{3-\mu^2}\,, 	
\end{eqnarray}
and $p, q, r, s$ are tetrahedral coordinates,
\begin{eqnarray}
p	&:=&	w-z-\sqrt{2}x	\,,\\
q	&:=&	w-z+\sqrt{2}x	\,,\\
r	&:=&	w+z+\sqrt{2}y		\,,\\
s	&:=&	w+z-\sqrt{2}y\,.
\end{eqnarray}
In the projective space description, the frequency $w$ can be put to
$1$. Decisive geometrical properties of a Kummer surface depend on the
value of the parameter $\mu$. This parameter turns out to be cubic
in 20 of Bateman's 21 constitutive components $\kappa_{IJ}$, see
Sec.1.5 below. Originally, a Kummer surface is defined in a real
3-dimensional space, but extensions to 3 complex dimensions are
straightforward.

A modern image of a Kummer surface, due to Rocchini \cite{Rocchini},
and to be compared with Fig.1, looks as follows:

\includegraphics[width=8truecm,height=6.0truecm]
{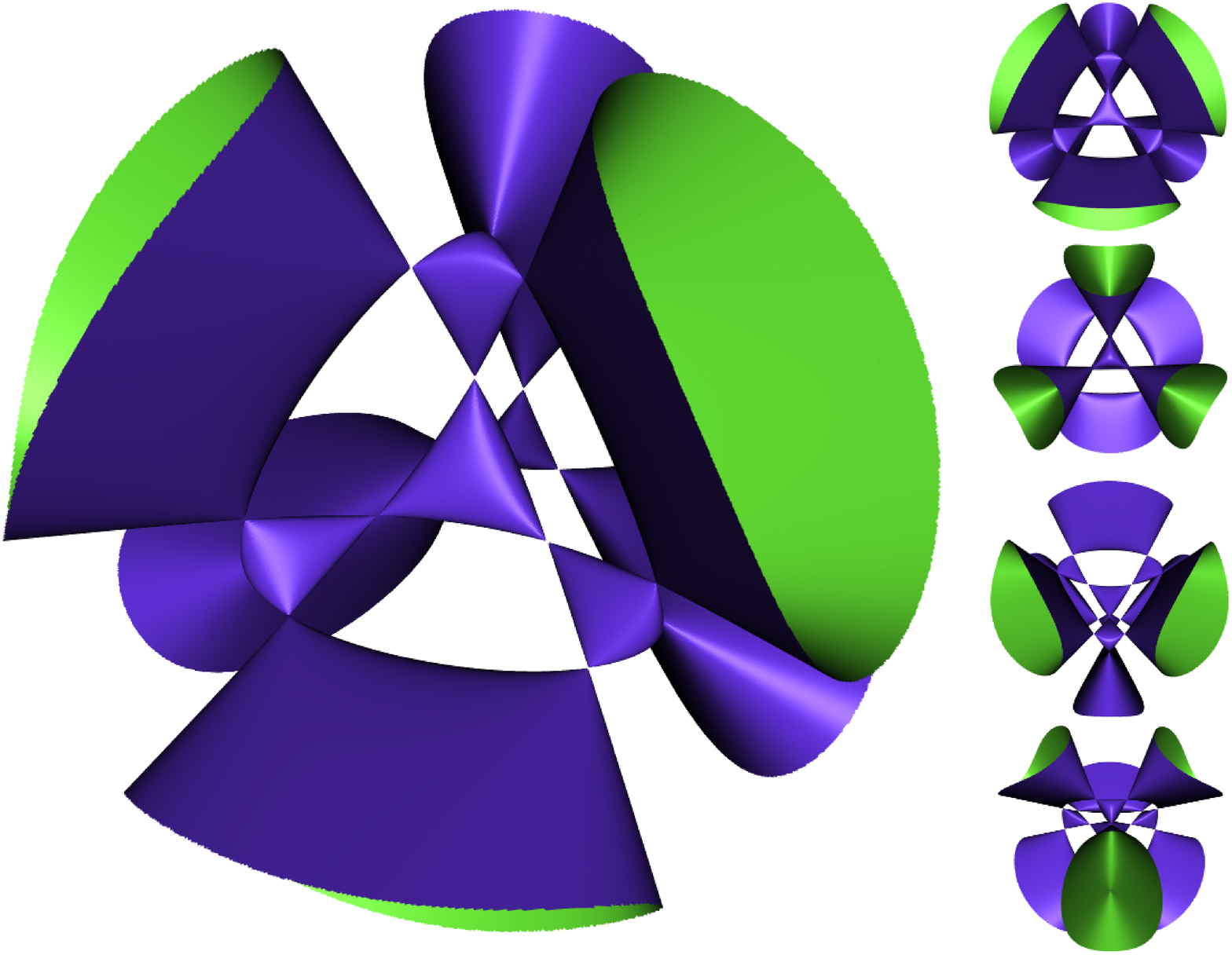}\\
Figure 2. {\it A Kummer surface due to Rocchini \cite{Rocchini}; for gypsum
and other models of Kummer surfaces, see Rowe \cite{Rowe}.}

We now consider the equality, for media with zero skewon piece, of
Fresnel and Kummer surfaces, which was first demonstrated by Bateman
\cite{Bateman_1910} in 1910. Electromagnetic waves have the property
that the fields $E_{a}$ and $B^{a}$ are orthogonal. The same holds
true for $H_{a}$ and $D^{a}$. In addition, the electric and magnetic
parts of the energy density are always equal. Even though the above
statements invoke the notion of orthogonality, a premetric
formulation is possible, see \cite{Birkbook}, p.126, as,
\begin{align}
  \epsilon^{ijkl}F_{ij}F_{kl}&=0 &\Leftrightarrow& &E_{a}B^{a}&=0,
  \label{geomopt1}\\\epsilon_{ijkl}{\cal H}^{ij}{\cal H}^{kl}&=0 &
  \Leftrightarrow& &H_{a}D^{a}&=0,\label{geomopt2}\\
  {\cal H}^{kl}F_{kl}&=0&\Leftrightarrow&
  &E_{a}D^{a}&=H_{a}B^{a}.\label{geomopt3}
\end{align}
The tensor density $\epsilon^{ijkl}=\pm 1, 0$ is the contravariant
analog of $\epsilon_{ijkl}$. Eqs.\eqref{geomopt1}--\eqref{geomopt3}
define geometric optics itself. For the general local and linear
electromagnetic response \eqref{const}, they become:
\begin{align}
  \epsilon^{IJ}F_{I}F_{J}&=0, \label{geomoptmed1}\\ 
\epsilon_{MN}\chi^{MI}\chi^{NJ}F_{I}F_{J}&=0,\label{geomoptmed2}\\
  \chi^{IJ}F_{I}F_{J}&=0,\label{geomoptmed3}
\end{align}
where $I,J,M,N=1,\dots,6$ denote pairs of antisymmetric indices. If
the medium has a vanishing skewon part,
\eqref{geomoptmed1}-\eqref{geomoptmed3} can be attributed a meaning in
projective geometry. The first equation dictates that $F_{ij}$
corresponds to a line in ${R}P^{3}$. As a matter of fact, one
can recognize that \eqref{geomoptmed1} identifies, effectively, the
Klein quadric. The antisymmetric $(0,2)$-tensors that obey
$\chi^{ijkl}F_{ij}F_{kl}=0$ and $\epsilon^{ijkl}F_{ij}F_{kl}=0$
correspond, in ${R}P^{3}$, to a quadratic line complex. As a
matter of fact, the electromagnetic response tensor density
$\chi^{ijkl}$ specifies a quadric, just like $\epsilon^{ijkl}$
specifies the Klein quadric. Finally, imposing \eqref{geomoptmed2} as
well as \eqref{geomoptmed1} and \eqref{geomoptmed3} entails that
$F_{ij}$ corresponds, in ${R}P^{3}$, to a singular line of the
quadratic line complex. These direct links between geometric optics
and projective geometry are explained in Bateman's article
\cite{Bateman_1910}. Nonetheless, the reader may want to consult the
papers
\cite{Delphenich:premetric,Delphenich:projectiveI,Delphenich:projectiveII}
by Delphenich, for a modern treatment.

In ${R}P^{3}$, the singular lines of the quadratic line complex
are tangent to the Kummer surface, as verified by Jessop
\cite{Jessop}. Remarkably, the Fresnel surface is also a result of
equations \eqref{geomoptmed1}--\eqref{geomoptmed3}. From this, it is
possible to deduce that, for the local and linear media with
${^{(2)}}\chi^{ijkl}=0$, the two surfaces coincide
\cite{Bateman_1910}.

Alternatively, one can prove such equality as follows, cf.\ Ruse
\cite{ruse1944sets, Ruse_1944b}. Equation \eqref{geomoptmed1} is
solved by $F_{ij}=2q_{[i}a_{j]}$. The quantities $q_{i}$ and $a_{i}$
turn out to be the 4-dimensional wave and polarization covectors. A
trivial substitution into \eqref{geomoptmed3}
yields:
\begin{gather}
  {\cal W}_{q}^{ij}\hspace{-1pt}[\chi]a_{i}a_{j}=0,\label{compcone}\\
 {\cal W}_{q}^{ij}\hspace{-1pt}[\chi]:=\chi^{ikjl}q_{k}q_{l}.\label{dispersionten}
\end{gather}
One recalls that $q_{i}$ and $a_{i}$ correspond to points in the real
projective space. Hence, \eqref{compcone} states that ${\cal W}_{q}^{ij}$
determines a complex cone in ${R}P^{3}$ for $a_{i}$ at $q_{i}$. To be
precise, only the symmetric part of the tensor \eqref{dispersionten}
contributes to \eqref{compcone}. Accordingly, it is useful to pick a
vector basis $\{e_{\bar{0}},\dots,e_{\bar{3}}\}$ such that
${\cal W}_{q}^{(\bar{i}\bar{j})}=\mbox{diag}(\lambda_{0},\dots,\lambda_{3})$,
with real eigenvalues. This allows one to reformulate \eqref{compcone}
as:
\begin{equation}
  \lambda_{0}a_{\bar{0}}^{2}+\lambda_{1}a_{\bar{1}}^{2}+\lambda_{2}a_{\bar{2}}^{2}
  +\lambda_{3}a_{\bar{3}}^{2}=0.\label{compconediag}
\end{equation}
Notably, the Kummer surface is the locus of points $q_{i}$ in $RP^{3}$
such that the complex cone \eqref{compconediag} simplifies to two
planes. By inspecting a table of the quadratic surfaces in
3-dimensions, one deduces that such factorization takes place
exclusively if ${\cal W}_{q}^{(ij)}$ has two vanishing eigenvalues. For
instance, when $\lambda_{0}=\lambda_{1}=0$, $\lambda_{2}<0$ and
$\lambda_{3}>0$, the complex cone is equal to two real intersecting
planes. Moreover, if $\lambda_{0}=\lambda_{1}=0$, $\lambda_{2}>0$ and
$\lambda_{3}>0$, equation \eqref{compconediag} specifies two imaginary
intersecting planes. All other cases are discussed in Section 4.18 of
the manual \cite{Zwillinger}. Demanding that ${\cal W}_{q}^{(ij)}$ has two
zero eigenvalues implies that its rank must be $0,1$ or $2$. If the
electromagnetic response has zero skewon part, as we assumed, the
tensor \eqref{dispersionten} is symmetric. So, when $q_{i}$
corresponds, in ${R}P^{3}$, to a point on the Kummer surface, the rank
of ${\cal W}_{q}^{ij}$ is $0,1$ or $2$. Analogously, $q_{i}$ belongs to the
Fresnel surface if and only if ${\cal W}_{q}^{ij}$ has rank less than or
equal to two \cite{Birkbook, Lindell:2005, Itin:2009aa}. In
consequence, Kummer surfaces are also Fresnel surfaces, and
vice-versa.

Before we collect our results, let us quote the precise formulation
from Avritzer and Lange \cite{Avritzer:2007}: ``\textit{It is
  well-known that the singular surface of a generic quadratic complex
  is a Kummer surface, i.e.\ a quartic surface in $\mathbb{P}^{3}$,
  smooth apart from 16 ordinary double points. Moreover every Kummer
  surface appears as the singular surface of a generic quadratic
  complex.}''

Summing up (see also \cite{FavaroHehl:2012}): the Kummer surface
emerged during studies of the light propagation in local and linear
media. Thus, light stood at the cradle of the Kummer surface. The
Fresnel surface of an electromagnetic response tensor density
$\chi^{ijkl}$, whose skewon part vanishes,
$\chi^{ijkl}-\chi^{klij}=0$, is a Kummer surface. Moreover, every
Kummer surface appears as the Fresnel surface of a generic
electromagnetic response tensor density, whose skewon part vanishes.

\subsection{Kummer surface and gravity}

Yes, Kummer started from optical considerations, more exactly from the
propagation of electromagnetic disturbances in the geometric optics
limit. Why should then Kummer's considerations be relevant for
gravity?

Well, both the electromagnetic and the gravitational fields are
massless, that is, they propagate with the speed of light, and their
waves are transversal with helicity 1 and helicity 2, respectively.
However, electromagnetic waves solve the {\it linear} Maxwell
equations, whereas gravitational waves obey the {\it
    nonlinear} Einstein equation. Of course, the Einstein
    equation is still linear in its highest (2\textsuperscript{nd})
    order derivatives.

    Let us recall how the TR-tensor density emerges in {\it
      electrodynamics.} We substitute the local and linear
    constitutive law ${\cal H}^{ij}=\frac{1}{2}\chi^{ijkl}F_{kl}$ into
    the source-free inhomogeneous Maxwell equation $\partial_j{\cal
      H}^{ij}=0$. Moreover, we express the field strength in terms of
    the potential, $F_{kl}=2\partial_{[k}A_{l]}$, and achieve
\begin{align}\label{goa}
  \partial_j\!\left(\chi^{ijkl}\partial_k A_l \right)=0& &\Rightarrow&
  &q_j\!\left(\chi^{ijkl}q_k a_l\right)=0\,
\end{align}
with $q_i\neq 0$ being the wave-covector.  Here, the geometric optics
limit is shown immediately -- see \cite{Birkbook} for details. By
introducing the characteristic matrix \eqref{dispersionten}, one can
rewrite \eqref{goa} as a homogeneous system of linear
equations,\footnote{The name \textit{characteristic matrix}, for the
  tensor \eqref{dispersionten}, was used in the paper
  \cite{Itin:2009aa}. }
\begin{equation}\label{linsyst}
 {\cal W}_{q}^{ij}\hspace{-1pt}[\chi]a_j=0\,.
\end{equation}
The field strength as determined in the geometric optics limit,
$F_{ij}=2q_{[i}a_{j]}$, is non-zero iff \eqref{linsyst} has two
linearly independent solutions, or more. We are thus prompted to
identify the wave-covectors such that the rank of ${\cal W}_{q}^{ij}$
is lower than 3. The search for all $q_{i}\neq 0$ that have this
property leads to the Fresnel equation (\ref{Fresnel}). Next,
\eqref{G4} specifies the Tamm-Rubilar and eventually, though not
uniquely, the Kummer tensor density (Sec.1.3, last paragraph).

Before we turn to the {gravitational} case in more detail, let us
first fix our corresponding notation. Our conventions are taken from
\cite{PRs}, unless stated otherwise. We work in the 4-dimensional {\it
  Riemann-Cartan} (RC) {\it spacetime} of the Poincar\'e gauge theory
of gravitation (see \cite{Reader}), a generalization of GR. Such a
spacetime is endowed with torsion $T_{ij}{}^k$ and curvature
$R_{ijk}{}^l$. If the torsion vanishes, we recover the Riemannian
spacetime, and the Riemannian quantities are denoted by a tilde
$\;\widetilde{\null}\;$, see also Obukhov \cite{Obukhov:2006gea}.

In {\it general relativity} (GR), the field strength, representing the
tidal forces, is the Riemann curvature tensor,\footnote{Often in GR,
  the $\widetilde{\Gamma}_{ij}{}^{k}$'s are named as ``gravitational
  field strength''. However, since they are non-tensorial, it seems
  more appropriate to us to call instead the tidal forces by this
  name.}
\begin{gather}\label{riemann}
  \widetilde{R}_{ijk}{}^{l}:=2(\partial_{[i}\widetilde{\Gamma}_{j]k}{}^{l}+
  \widetilde{\Gamma}_{[i|m}{}^l\,\widetilde{\Gamma}_{|j]k}{}^m)\,,\\
  \text{with}\quad \widetilde{\Gamma}_{ij}{}^{k}:=\frac
  12g^{kl}(\partial_i g_{jl}+\partial_j g_{il}-\partial_l g_{ij})\,.
\end{gather}
Here, the Riemann (or Levi-Civita) connection $
\widetilde{\Gamma}_{ij}{}^{k}$ is defined in terms of the
gravitational potential, the metric $g_{ij}$.

Owing to the symmetries $\widetilde{R}^{(ij)kl}=
\widetilde{R}^{ij(kl)}=0$, $ \widetilde{R}^{ijkl}=
\widetilde{R}^{klij}$ and $\widetilde{R}^{[ijkl]}=0$, the Riemann
curvature tensor has 20 independent components. We separate these into
10 independent entries for the Ricci tensor
$\text{$\widetilde{\text{R}}$ic}_{ij}:= \widetilde{R}_{kij}{}^k$, and
10 independent entries for the Weyl tensor $\widetilde{C}^{ijkl}$. The
latter, also known as the conformal Weyl curvature, is the trace-free
part. Contracting the Ricci tensor with the inverse metric leads to
the curvature scalar,
$\widetilde{R}:=g^{ij}\text{$\widetilde{\text{R}}$ic}_{ij}$. We remove
this single independent component from the Ricci tensor, and obtain
the trace-free part ``Ricsymf\hspace{1pt}".  In conclusion, see
\cite{PRs},
\begin{align}
  \mbox{Riem} &= \mbox{Weyl} \oplus \mbox{Ricsymf} \oplus \mbox{Scalar}\,.\\
  \nonumber 20 &=\hspace{7.05pt}10\hspace{7.05pt}\oplus
  \hspace{17.7pt} 9\hspace{17.7pt}\oplus 1\,.
\end{align}

This was the kinematics of the gravitational field strength
$\widetilde{R}^{ijkl}$. Let us now turn to the dynamics: the Einstein
field equation without cosmological constant and sources,
\begin{equation}\label{Einstein}
  \text{$\widetilde{\text{R}}$ic}_{ij}-\frac 12 g_{ij}\widetilde{R}=0\qquad
  \text{or}\qquad \text{$\widetilde{\text{R}}$ic}_{ij}=0\,,
\end{equation}
determines that, in vacuum, only the Weyl tensor
$\widetilde{C}^{ijkl}$ is left over for describing the gravitational
field strength. The {\it Petrov classification} (1954) of the
gravitational vacuum field, see Penrose et al.\ \cite{Penrose} and
Stephani et al.\ \cite{S_K_2004}, is based on the algebraic properties
of the Weyl tensor alone. One way is to study, for the bivector
$X^{ij}=-X^{ji}$, the eigenvalue equation
\begin{equation}\label{Petrov}
\frac 12 \widetilde{C}^{ij}{}_{kl}X^{kl}=\lambda X^{ij}\,.
\end{equation}
Another, equivalent way is to determine the {\it principal null
  directions} of the Weyl tensor according to \cite{S_K_2004}
\begin{equation}\label{pnd}
  \xi^{[i}\bigl(\xi_{m}\widetilde{C}^{j]mn[k}\xi_{n}\bigr)\xi^{l]}=0\,,\qquad
  \text{with}\qquad g^{ij}\xi_{i}\xi_{j}=0.
\end{equation}
So, the covector $\xi_i\neq 0$ is explicitly required to be null
(light-like). The 2nd rank tensor occurring inside the
parentheses of (\ref{pnd}),
\begin{equation}\label{2ndranktensor}
  W_\xi^{ij}\hspace{-1pt}[\widetilde{C}]:=\widetilde{C}^{ikjl}\xi_k\xi_l\,,
\end{equation}
is the analog of ${\cal W}_q^{ij}\hspace{-1pt}[\chi]$, see
(\ref{dispersionten}). Thus, the Weyl tensor $\widetilde{C}^{ikjl}$
mimics a corresponding electromagnetic response tensor $\chi^{ikjl}$,
and there is small wonder that demanding $W_\xi^{ij}\hspace{-1pt}
[\widetilde{C}]$ to have rank $0,1$ or $2$ yields a Kummer tensor
$K^{ijkl}[\widetilde{C}]$.

In this analogy, we compare electromagnetic wave propagation in local
and linear media, that is, Maxwell's linear field theory {\it in
  matter,} with the nonlinear gravitational wave propagation {\it in
  vacuum}, that is, with the nonlinear Einstein equation in free
space.

But this analogy teaches us one more thing: The electromagnetic
response tensor $\chi^{ijkl}$, only in the simplest nontrivial
case, depends on 20 independent components, see (\ref{chidec}) for
${}^{(2)}\chi^{ijkl}={}^{(3)}\chi^{ijkl}=0$. Can we generalize the 20
components' Riemann curvature $\widetilde{R}_{ijk}{}^l$ to an object
with 36 independent components? Yes, we can! In the gauge theory of
the Poincar\'e group, see \cite{Reader}, Part B, the field strength,
besides Cartan's torsion, is the curvature tensor of a Riemann-Cartan
space, 
\begin{equation}\label{RCcurv}
  {R}_{ijk}{}^{l}:=2(\partial_{[i}{\Gamma}_{j]k}{}^{l}+
  {\Gamma}_{[i|m}{}^l\,{\Gamma}_{|j]k}{}^m)\,,
\end{equation}
with 36 independent components, in strict analogy to
$\chi^{ijkl}$. The Riemann-Cartan connection $\Gamma_{ij}{}^k$,
entering the definition (\ref{RCcurv}), is metric-compatible and
carries the torsion tensor $T_{ij}{}^k:=2\Gamma_{[ij]}{}^k$. The Weyl
tensor of a Riemann-Cartan space $C^{ijkl}$ has, like its
Riemannian counterpart, 10 independent components and is traceless.

In GR, starting with the curvature $\widetilde{R}_{ijk}{}^{l}$, we can
construct the Kummer tensor $K^{ijkl}[\widetilde{R}]$. It is possible
to show, see Ruse \cite{Ruse_1944b}, that the surface generated by
$K^{ijkl}[\widetilde{R}]\,\xi_{i} \xi_{j}\xi_{k}\xi_{l}=0$ is a Kummer
surface.\footnote{Incidentally, the nodes of the Kummer surface can be
  expressed by $K^{ijkl}[\widetilde{R}]\, \xi_{i} \xi_{j}\xi_{k}= 0$,
  see Ruse \cite{ruse1944sets}.} Let us now generalize to a
Riemann-Cartan spacetime, having curvature tensor $R_{ijk}{}^{l}$. It
is an open problem whether
\begin{equation}\label{maynotkummer}
K^{ijkl}[R]\,\xi_{i}\xi_{j}\xi_{k}\xi_{l}=0
\end{equation} 
always leads to a Kummer surface, since it may occur that
$R^{ijkl}\neq R^{klij}$. As a matter of fact, this is algebraically
the same as having a medium with a finite skewon piece (drawing the
analogy involves raising three indices of the curvature tensor with
the metric $g^{ij}$). Nevertheless, \eqref{maynotkummer} does result
in a surface that is equivalent to a Fresnel surface, and which
encodes information about the principal null directions of
$R_{ijk}{}^{l}$.

We define the principal null directions of a curvature tensor in
Riemann-Cartan spacetime as the covectors such that
\begin{gather}
  \xi^{[i}\bigl(\xi_{m}R^{j]mn[k}\xi_{n}\bigr)\xi^{l]}=0,\label{principal}\\
g^{ij}\xi_{i}\xi_{j}=0. \label{null}
\end{gather}
In particular, \eqref{principal} is obtained by generalizing
(\ref{pnd}) from Riemannian to Riemann-Cartan spacetimes. As an aside,
we were able to prove that, if $R^{ikjl}\xi_{k}\xi_{l}\neq 0$,
\eqref{principal} implies \eqref{null}. One could then hope that all
covectors that satisfy \eqref{principal} are null, whence the
requirement \eqref{null} could be dropped. This hope is, in fact,
untenable. Let us consider the case where $R^{ijkl}=\alpha
\epsilon^{ijkl}$, so that $R^{ikjl}\xi_{k}\xi_{l}=0$. Then, the
requirement \eqref{principal} is fulfilled by an arbitrary covector,
which may not be null. One then concludes that \eqref{null} must be
kept as an explicit requirement.

The following observation motivated, to a good extent, our interest in
Kummer surfaces: the principal null directions of
$R_{ijk}{}^{l}$ belong to the intersection of the
hypersurface generated by \eqref{maynotkummer} and the light-cone. The
key is to demonstrate that, because of \eqref{principal}, the covector
$\xi_{i}$ fulfills \eqref{maynotkummer}. In addition, \eqref{null}
states that $\xi_{i}$ must also belong to the light-cone.

Let us look at this proof more in detail. The first step is to define
an equivalent of the tensor \eqref{2ndranktensor}, namely
$W_{\xi}^{ij}\hspace{-1pt}[R]:=R^{ikjl}\xi_{k}\xi_{l}$. One can then rewrite
\eqref{principal} as
\begin{equation}\label{principalw}
  \xi^{[i}W_{\!\xi}\vphantom{\xi}^{j][k}\xi^{l]}=0.
\end{equation}
Decomposing \eqref{principalw} in space+time yields a set of four simultaneous equations
\begin{equation}
\begin{cases}\label{principalwdec}
  \xi^{[0}W_{\!\xi}\vphantom{\xi}^{a][0}\xi^{b]}&=0,\\
  \xi^{[0}W_{\!\xi}\vphantom{\xi}^{a][b}\xi^{c]}&=0,\\
  \xi^{[a}W_{\!\xi}\vphantom{\xi}^{b][0}\xi^{c]}&=0,\\
  \xi^{[a}W_{\!\xi}\vphantom{\xi}^{b][c}\xi^{d]}&=0.\\
\end{cases}
\end{equation}
The indices $a,b,c,d$ range from $1$ to $3$. Next, one considers a
vector basis $\{e_{\bar{0}},e_{\bar{a}}\}$ such that $\xi^{\bar{0}}=1$
and $\xi^{\bar{a}}=0$. In this vector basis, the second, third and
fourth equations of \eqref{principalwdec} are trivially fulfilled. The
condition \eqref{principalw} is then equivalent to the first equation
of the set, which takes the form
\begin{equation}
  W_{\!\xi}^{\bar{a}\bar{b}}=0.
\end{equation}
By contrast, $W_{\!\xi}^{\bar{0}\bar{0}}, W_{\!\xi}^{\bar{0}\bar{a}}$ and $W_{\!\xi}^{\bar{a}\bar{0}}$ remain arbitrary. It follows that, using the tensor product $\otimes$,
\begin{align}\label{principalwrem}
  W_{\!\xi}&=W_{\!\xi}^{\bar{0}\bar{0}} e_{\bar{0}}\otimes e_{\bar{0}}
  +W_{\!\xi}^{\bar{0}\bar{a}}e_{\bar{0}}\otimes
  e_{\bar{a}}+W_{\!\xi}^{\bar{b}\bar{0}}e_{\bar{b}}\otimes e_{\bar{0}}\nonumber\\
  &=e_{\bar{0}}\otimes(W_{\!\xi}^{\bar{0}\bar{0}}e_{\bar{0}}+W_{\!\xi}^{\bar{0}\bar{a}}
e_{\bar{a}})+W_{\!\xi}^{\bar{b}\bar{0}}e_{\bar{b}}\otimes e_{\bar{0}}.
\end{align}
When one defines the vectors
$S_{\xi}:=W_{\!\xi}^{\bar{0}\bar{0}}e_{\bar{0}}+W_{\!\xi}^{\bar{0}\bar{a}}e_{\bar{a}}$
and $T_{\xi}:=W_{\!\xi}^{\bar{b}\bar{0}}e_{\bar{b}}$,
\eqref{principalwrem} translates into $W_{\!\xi}=e_{\bar{0}}\otimes
S_{\xi}+T_{\xi}\otimes e_{\bar{0}}$. In an arbitrary 4-dimensional
vector basis $\{e_{i}\}$, this reads
\begin{equation}
  W_{\!\xi}^{ij}=e_{\bar{0}}^{\ i}S_{\xi}^{j}+T_{\xi}^{i}e_{\bar{0}}^{\ j},
\end{equation}
where $e_{\bar{0}}=e_{\bar{0}}^{\ i}e_{i}$. One recalls that $e_{\bar{0}}^{\ i}=\xi^{i}$, and finds that \eqref{principal} is equivalent \nolinebreak to
\begin{equation}
   W_{\!\xi}^{ij}=\xi^{i}S_{\xi}^{j}+T_{\xi}^{i}\xi^{j},
\end{equation}
for some $S_{\xi}^{i}$ and $T_{\xi}^{i}$. Hence, the condition
\eqref{principal} implies that $W_{\xi}^{ij}$ has rank $0,1$ or
$2$. Let us momentarily leave this result to one side, and proceed on
a different front.  As we mentioned before, the surface generated by
\eqref{maynotkummer} is algebraically the same as a Fresnel
surface. Consequently, one can exploit the results of
\cite{Birkbook,Lindell:2005,Itin:2009aa} in geometric optics, and
deduce that $\xi_{i}$ belongs to the hypersurface generated by
\eqref{maynotkummer} if and only if the rank of $W_{\!\xi}^{ij}$ is
$0,1$ or $2$. Bringing it all together, when $\xi_{i}$ fulfills
\eqref{principal}, it also satisfies \eqref{maynotkummer}. At the same
time, $\xi_{i}$ is required to be null \eqref{null}. Therefore, the
principal null directions of $R_{ijk}{}^{l}$ belong to the
intersection of the hypersurface generated by \eqref{maynotkummer} and
the light-cone. If $R^{ijkl}=R^{klij}$, one can further conclude that
the hypersurface describes, in ${R}P^{3}$, a Kummer surface.

Let us observe that one may classify the principal null directions of
$R_{ijk}{}^{l}$ according to the rank of $W_{\xi}^{ij}\hspace{-1pt}[R]$.

\subsection{Outline, conventions}

In Sec.~3, we {\it decompose} the premetric Kummer tensor density
${\cal K}^{ijkl}[{\cal T}]$, which has 136 independent components,
into smaller pieces. We determine its irreducible pieces under the
linear group $GL(4,R)$, see Eq.(\ref{irr-canonical}). Especially, we
also spell out the irreducible decomposition of ${\cal K}^{ijkl}[{\cal
  T}]$, if the latter is expressed only in terms of one single
irreducible piece of ${\cal T}^{ijkl}$, see Figure 5.

In Sec.~4, we assume the presence of a spacetime {\it metric} that
allows to contract the Kummer tensor once or twice. Accordingly, we
extract the completely trace-free part of the Kummer tensor and,
moreover, we decompose ${\cal K}$ under the Lorentz group
$SO(1,3)$. As a special case, this yields the $SO(1,3)$ decomposition
of the Tamm-Rubilar tensor into 3 pieces according to $35=25+9+1$,
which is the new result.

%
%
Our conventions of physics are basically taken from Landau-Lifshitz
\cite{LL}: (holonomic) coordinates are denoted by Latin indices
$i,j,k,...=0,1,2,3$, spatial coordinate indices by $a,b,...=1,2,3$,
(anholonomic) frames (tetrads) by Greek indices
$\a,\b,\g,...=0,1,2,3$. Our metric has signature $(-+++)$. Our
conventions in differential geometry are taken from Schouten
\cite{Schouten_1954}, see also Penrose \& Rindler \cite{Penrose}. Only
our torsion $T_{ij}^{k}=2S_{ij}^{k}|_{\text{Schouten}}$.  Symmetrizing
of indices is denoted by parentheses, $(ij):=\{ij+ji\}/2!$,
antisymmetrization by brackets $[ij]:=\{ij+ji\}/2!$, with
corresponding generalizations $(ijk):=\{ijk+jik+kij+\cdots\}/3!$,
etc.; indices standing between two vertical strokes ${|\hspace{5pt}|}$
are excluded from the (anti)symmetrization process.

Our conventions in exterior calculus are displayed in
\cite{PRs,Birkbook}. The Hodge star is denoted by $^\star$, the dual
with respect to the Lie-algebra indices by $^{(*)}$. In tensor
analysis, it is not necessary to distinguish between the Hodge dual
and the Lie dual: we simply use $^{*}$. The totally antisymmetric
Levi-Civita symbol (a tensor density) is $\epsilon_{ijkl}=\pm1,0$. The
``diamond'' dual $^\diamond$, see \cite{Birkbook} (sometimes called Poincar\' e
dual, see \cite{Delphenich:premetric}), is built with the premetric
$\epsilon_{ijkl}$.

\section{Kummer tensor density ${\cal K}^{ijkl}[{\cal T}]$ 
in spacetime newly defined}

\subsection{Its definition in terms of a doubly antisymmetric 4th rank
tensor density ${\cal T}^{ijkl}$}

Consider an arbitrary fourth rank tensor density of type
$\left(\begin{smallmatrix}4\\0\end{smallmatrix}\right)$ that is
antisymmetric in its first and its last pair of indices:
\begin{equation}\label{symmetries}
  {\cal T}^{ijkl}\qquad\text{with}\qquad{\cal T}^{(ij)kl}=0\,\quad\text{and}\quad
    {\cal T}^{ij(kl)}=0\,.
\end{equation}
Then, ${\cal T}$ can be thought of as a $6\times6$ matrix, which can
be decomposed into a symmetric traceless, an antisymmetric, and a
trace part. This corresponds, see the analogous case of $\chi$ in
(\ref{chidec}), to an irreducible decomposition of ${\cal T}^{ijkl}$
under the the action of the group $GL(4,R)$:\footnote{The {\it
    principal part} of ${\cal T}^{ijkl}$ can be put into the explicit form
\begin{eqnarray}\label{spec4} \nonumber 
  {}^{(1)}{\cal T}^{ijkl}& =&\frac 16\Big[2\left({\cal T}^{ijkl}
    +{\cal T}^{klij}\right)-\left({\cal T}^{iklj}+{\cal T}^{ljik}\right)
-\left({\cal T}^{iljk}+{\cal T}^{jkil}\right)\Big]\,.
\end{eqnarray}}
\begin{eqnarray}\label{Tdec}
  {\cal T}^{ijkl}&=&\,^{(1)}{\cal T}^{ijkl}+
  \,^{(2)}{\cal T}^{ijkl}+
  \,^{(3)}{\cal T}^{ijkl}\,.\\ \nonumber 36
  &=&\hspace{11pt} 20\hspace{15pt}\oplus \hspace{13pt}15\hspace{13pt}
  \oplus \hspace{21pt}1\,.
\end{eqnarray}

Motivated by the existence, see (\ref{G4}), of the Tamm-Rubilar tensor
density ${\cal G}^{ijkl}[\chi]$ of classical electrodynamics, and
starting from (\ref{G4doublediamond}), we want to define, as a
prototype, the Kummer tensor density ${\cal K}^{ijkl}[{\cal T}]$, which
is {\it cubic} in the fourth rank tensor density $\cal T$ of
equation (\ref{symmetries}).

As long as one has no metric available---we call this the {\it
  premetric} situation---one can
only use the totally antisymmetric Levi-Civita symbol
$\epsilon_{ijkl}=\pm 1,0$ for lowering indices. We define, see above,
the ``diamond'' (single) dual by
\begin{equation}\label{singlediamond}
{\cal T}^{\diamond\, ij}{}_{kl}=\frac 12 {\cal T}^{ijcd}\epsilon_{cdkl}
\end{equation}
and the double dual by
\begin{equation}\label{doublediamond}
{^\diamond}{\cal T}^{\,\diamond}_{\hspace{2pt}ijkl}=\frac 12  \epsilon_{ijab}
{\cal T}^{\diamond\, ab}{}_{kl} =\frac 14 \epsilon_{ijab}{\cal T}^{abcd}
\epsilon_{cdkl}\,.
\end{equation}
Clearly, we have once again the symmetries $^{\diamond}{\cal
  T}^\diamond{}_{(ij)kl}= {}^{\diamond}{\cal
  T}^\diamond{}_{ij(kl)}=0$, and this double dual can also be
decomposed according to the scheme (\ref{Tdec}).

Keeping in mind the definition (\ref{G4doublediamond}) and
desymmetrizing it, let us now introduce the prototype of a Kummer tensor
density that corresponds to the fourth rank tensor density $\cal T$ of
equation (\ref{symmetries}):
\begin{eqnarray}\label{KummerDef}
\label{KummerT'}
\boxed{{{\cal K}^{ijkl}[{\cal T}]}:= {\cal T}^{ai bj}{}\,
{^\diamond}{\cal T}^\diamond_{\hspace{2pt}acbd}{\cal T}^{ckdl}\,.}
\end{eqnarray}

If we switch the index pairs according to ${{\cal K}^{klij}[{\cal
    T}]}:= {\cal T}^{ak bl}{}\, ^{\diamond}{\cal
  T}^\diamond{}_{\hspace{-2pt}acbd}{\cal T}^{cidj}={\cal T}^{cidj}\,
^{\diamond}{\cal T}^\diamond{}_{\hspace{-2pt}acbd} {\cal T}^{ak
  bl}{}$, then, by renaming the indices $c,d$ and $a,b$, we
immediately recognize the symmetry
\begin{equation}\label{paircom}  \boxed{ {{\cal K}^{ijkl}[{\cal T}]}={{\cal
    K}^{klij}[{\cal T}]}\,.}
\end{equation}
No other algebraic symmetries of the Kummer tensor density are known,
provided $\cal T$ carries no additional symmetries beyond
(\ref{symmetries}). Thus, we can think of ${{\cal K}^{ij|kl}[{\cal
    T}]}$ as a 16$\times$16 matrix. Because of (\ref{paircom}), it is
a symmetric $16\times16$ matrix {\it with 136 independent components.}

A Kummer tensor belonging to the Riemann curvature tensor
$\widetilde{R}^{ijkl}$ of general relativity was defined by Zund
\cite{Zund_1969} explicitly in his Eq.(16) and earlier by Ruse
\cite{Ruse_1944b} implicitly in his Eqs.(5.15) and (5.7), see also
\cite{Ruse_1936,ruse1944sets}. Our convention for the position of the
indices of the Kummer tensor density turned out to be in harmony with
that of Ruse and Zund. Note, however, that they worked in a Riemannian
space with a Riemannian metric, whereas our definition is {\it
  premetric} and, as such, appreciably more general. In premetric
electrodynamics, this more general definition (\ref{KummerT'}) is
required, which we propose here {\it for the first time.}

\subsection{In premetric electrodynamics}

In exterior calculus, the premetric form of Maxwell's equations reads,
\begin{eqnarray}\label{Maxwell}
                  d H = J\,, \qquad   d F = 0\,, 
\end{eqnarray}
with the excitation 2-form $H=(-{H_a},{D^b})$, the electric current
3-form $J=({-j^a},\rho)$, and the field strength 2-form
$F=({E_a},{B^b})$, see \cite{Birkbook}. Maxwell's equations, as
displayed in (\ref{Maxwell}), are manifestly invariant under
coordinate and frame transformations and are visibly independent of
the metric, since the $d$'s denote exterior ({\it not} exterior {\it
  covariant}) differentiation. With the formulas
\begin{eqnarray}\label{components}
  H&=&{\scriptstyle{\frac 12}} \epsilon_{ijkl}{\cal H}^{kl}dx^i\wedge dx^j\,,
  \qquad J= {\scriptstyle{\frac 16}}
  \epsilon_{ijkl}{\cal J}^l dx^i\wedge dx^j\wedge dx^k\,, \nonumber\\
 F&=&{\scriptstyle{\frac 12}} F_{ij}dx^i\wedge dx^j\,,     
\end{eqnarray}
we recover the coordinate component version (\ref{Maxwellcomp}) of
Maxwell's equations.

With the local and linear constitutive law (\ref{const}), ${\cal
  H}^{ij}=\frac 12\chi^{ijkl}F_{kl}$, one finds the Kummer tensor
density of premetric electrodynamics as
\begin{equation}\label{KummerMax}
{\cal K}^{ijkl}[{\chi]}= {\chi}^{ai bj}{}\,
{^\diamond}{\chi}^\diamond_{\hspace{2pt}acbd}{\chi}^{ckdl}\,.
\end{equation}
Thus, recalling how we arrived at ${\cal K}^{ijkl}[\chi]$ in the
context of (\ref{G4doublediamond}), we immediately retrieve the
TR-tensor density

\begin{center}
\includegraphics[width=8cm]{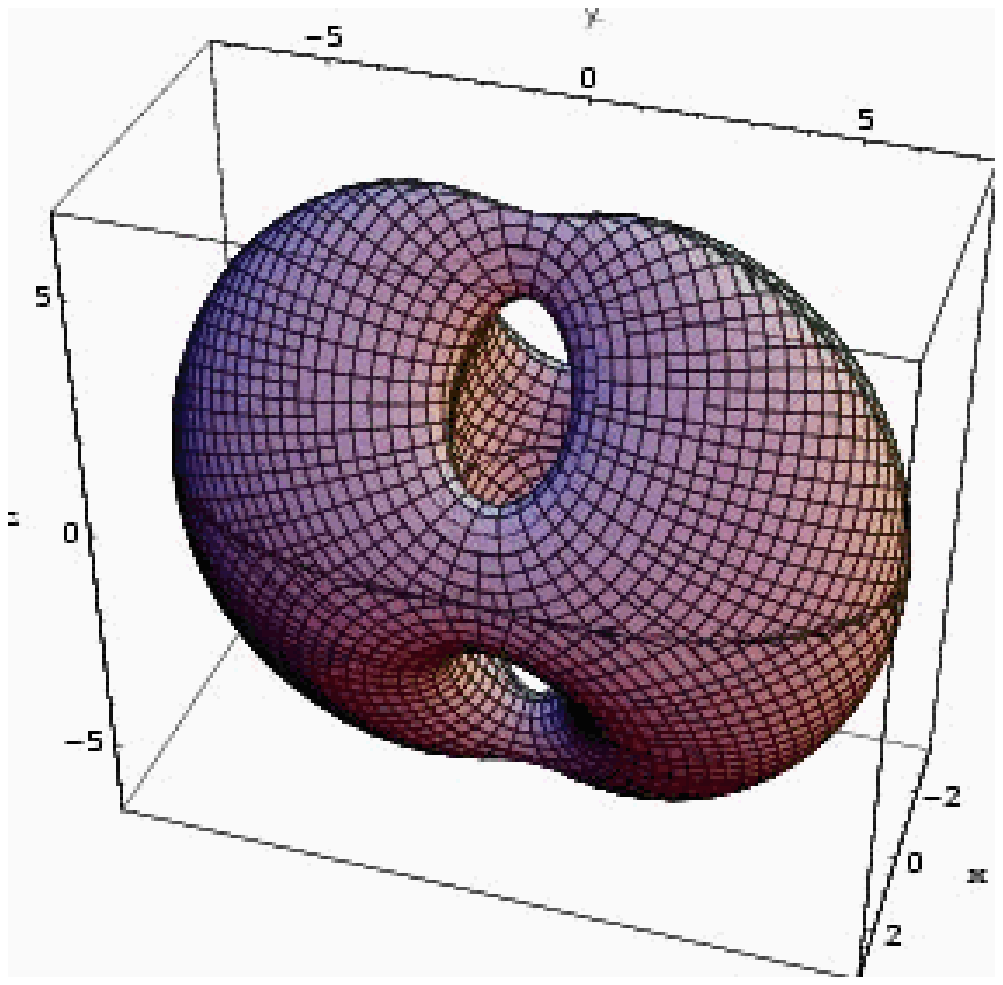}
\end{center}

Figure 3. {\it Generalized Fresnel wave surface for anisotropic
  permittivity, trivial impermeability, and with a {\it spatially
    isotropic skewon} piece. We use the dimensionless variables $x :=
  cq_1/q_0,\,y := cq_2/q_0,\,z := cq_3/q_0$, see
\cite{Hehl:2005hu}.}\bigskip

\begin{center}
\includegraphics[width=11cm]{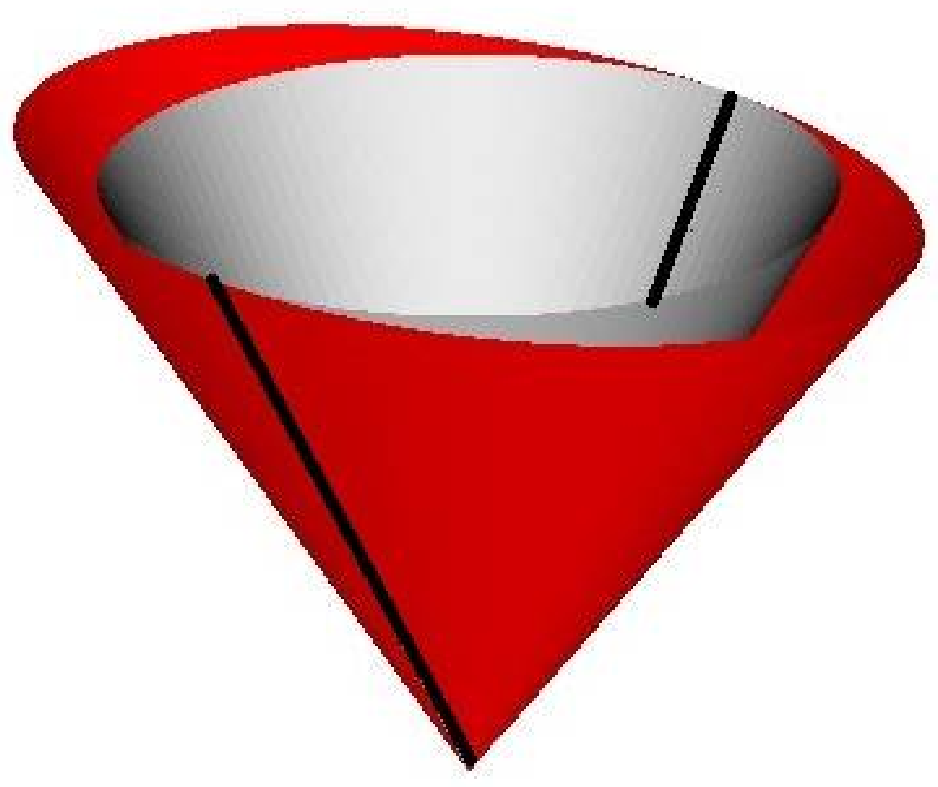}
\end{center}

\vspace{-26pt}Figure 4.  {\it {The two cones of the generalized Fresnel wave
    surface for trivial permittivity and impermeability, with a
    one-component skewon piece $\chi^{1213}$ (or ${\not\hspace{-3pt}
    S}_{01}$).  The axis of rotation is directed along the time
    coordinate.  The two optical axes are lying in the plane $q_2=0$,
    the fourth coordinate $q_3$ is suppressed. In contrast, double
    cone media {\em for vanishing skewon piece} have been investigated
    by Dahl and Favaro \cite{Dahl:2011he,DahlFavaro2014}.}}\bigskip

\begin{eqnarray}\label{TR21}
   {\cal G}^{ijkl}[\chi] = \frac{1}{6}\,{\cal K}^{(ijkl)}[\chi]\,.
\end{eqnarray}
Accordingly, the totally symmetric piece ${\cal K}^{(ijkl)}[\chi]$,
with its 35 independent components, can, up to a factor, be observed
in crystal optics. In Figures 3 and 4, we display generalized Fresnel
wave surfaces for media that carry a skewon piece, in order to convey
an idea of the numerous possible physical situations.

The TR-tensor density has an operational interpretation. This is not
yet the case for the other, less symmetric pieces of the Kummer tensor
density ${\cal K}^{ijkl}[\chi]$. It is one of the goals of our paper
to come to a better understanding of these new pieces, which are
beyond the TR-tensor density.

If we restrict ourselves to local and linear {\it reversible}
electrodynamics, that is, if the skewon piece vanishes,
$^{(2)}\chi^{ijkl}=0$, then we recover the case which was considered
by Bateman \cite{Bateman_1910}, see our equation (\ref{constBateman})
with $\kappa_{IJ}=\kappa_{JI}$. Then,
$\chi={}^{(1)}\!\chi+{}^{(3)}\!\chi$ has only $20+1$ independent
components. Exactly for this reversible case, Bateman, by using tools
of projective geometry, see Hudson \cite{Hudson_1903} and Jessop
\cite{Jessop}, has demonstrated that the generalized Fresnel wave
surfaces are {\it Kummer surfaces}. This fact justifies to call ${\cal
  K}^{ijkl}$ the Kummer tensor---at least for $^{(2)}\chi=0$. However,
we do not know whether the wave surfaces in Figs.\ 3 and 4 represent
Kummer surfaces or not, since they incorporate a non-vanishing
$^{(2)}\chi$.

\subsection{In general relativity (Riemann spacetime)}

Let us turn to the metric case: if a Riemannian metric with Lorentz
signature is prescribed, we can define the scalar density of weight
$+1$ as $\sqrt{-g}$, with $g:=\det g_{rs}$. Then the densities can be
transformed to tensors (see \cite{Birkbook}, p.218),
\begin{eqnarray}\label{dens}
  T^{ijkl}=(\sqrt{-g})^{-1} {\cal T}^{ijkl}\,,\quad 
  K^{ijkl}=(\sqrt{-g})^{-1} {\cal K}^{ijkl}\,,\quad  
\eta_{ijkl}:=\sqrt{-g}\,\epsilon_{ijkl}\,,
\end{eqnarray}
and the dual $^*:=\sqrt{-g}\;^\diamond$ is built with this unit
tensor. 

In general relativity (GR), $T^{ijkl}$ is identified with the Riemann
curvature tensor $\widetilde{R}^{ijkl}$, that is,
$T^{ijkl}=\widetilde{R}^{ijkl}$. We recall the algebraic symmetries of
the Riemann curvature tensor:
\begin{equation}\label{RiemannSym}
  \widetilde{R}^{(ij)kl}=0,\quad\widetilde{R}^{ij(kl)}=0;\qquad\widetilde{R}^{ijkl}=
  \widetilde{R}^{klij};\qquad\widetilde{R}^{[ijkl]}=0.
\end{equation}
The first two symmetries qualify $\widetilde{R}$ to be identified with
$T$, that is, we have $36$ independent components left. The pair
commutator symmetry corresponds, algebraically, to the vanishing of
the skewon piece in electrodynamics. Thus, $21$ independent components
are left over, in the ``reversible'' case. Eventually, the vanishing
of the totally antisymmetric piece (1 component) leaves us with $20$
independent components for the Riemann curvature
$\widetilde{R}^{ijkl}$. Electrodynamically speaking, only the
principal part of the curvature survives. In the parlance of, for
example, Gilkey \cite{Gilkey}, $^{(1)}\chi$ is an {\it
  algebraic} Riemann curvature tensor.

With the identification $T=\widetilde{R}$, the Kummer tensor density
becomes the plain gravitational Kummer tensor of Zund
\cite{Zund_1969},
\begin{eqnarray}
  \label{KummerT''}
  {{K}^{ijkl}[\widetilde{R}]}= \widetilde{R}^{ai bj}\,
  {^*}\!\widetilde
{R}^*_{\hspace{2pt}acbd}\,\widetilde{R}^{ckdl}\,.
\end{eqnarray}

Since $\widetilde{R}^{ijkl}$ exhibits only on 20 independent
components, its Kummer tensor possesses, besides the conventional
pairwise symmetry
\begin{equation} \label{convSym}
{{K}^{ijkl}[\widetilde{R}]}=  {{K}^{klij}[\widetilde{R}]}\,,
\end{equation}
also an additional algebraic symmetry, namely
\begin{equation}\label{addSymRiemann}
  K^{ijkl}[\widetilde{R}]=K^{jilk}[\widetilde{R}]\,.
\end{equation}
This can be shown directly, but later, in Eq.\ (\ref{spec20}), the
proof of (\ref{addSymRiemann}) will be much more illuminating.

Incidentally, to our knowledge, the Kummer tensor
${{K}^{ijkl}[\widetilde{R}]}$ of Zund \cite{Zund_1969} was forgotten
altogether. We could not find any new reference to it.

The curvature tensor of GR, $\widetilde{R}^{ijkl}$ (20 independent
components), can be irreducibly decomposed, under the local Lorentz
group, into the Weyl tensor $\widetilde{C}^{ijkl}$ (10 components),
the trace-free Ricci tensor $\widetilde{\text{R}}\text{ic}_{ij}-\frac 14
\widetilde{R}\,g_{ij}$ (9 components, with
$\widetilde{\text{R}}\text{ic}_{ij}:=\widetilde{R}_{kij}{}^k$),
and the curvature scalar
$\widetilde{R}:=\widetilde{\text{R}}\text{ic}_{k}{}^k$(1
component). 

Accordingly, we can define three different Kummer tensors for each of
these pieces. In practical applications, however, only the Kummer
tensor attached to the Weyl tensor, $K^{ijkl}[\widetilde{C}]$, will be
of importance. For a matter-free region of spacetime (vacuum), we
have, according to Einstein's field equation without cosmological
constant, $\widetilde{\text{R}}\text{ic}_{ij}=0$. In other words, a
{\it vacuum gravitational field} is described by a non-vanishing {\it
  Weyl tensor,} $\widetilde{C}^{ijkl}\ne 0$. In turn, its Kummer
tensor, $K^{ijkl}[\widetilde{C}]$, will be relevant in this
context. We will discuss this in Sec.2.5.

\subsection{In Poincar\'e gauge theory of gravity (Riemann-Cartan
  spacetime)}

In the RC-space, the curvature tensor $R_{ijk}{}^l$ has only the
following algebraic symmetries, 
\begin{equation}\label{SymmRC}
R^{(ij)kl}=0,\quad R^{ij(kl)}=0\,.
\end{equation}
Thus, it has 36 independent components, just like the electromagnetic
response tensor $\chi^{ijkl}$, which, accordingly, is called an {\it
  algebraic} RC-curvature tensor. The Kummer tensor of the RC-cuvature
reads
\begin{eqnarray}\label{KummerRC1}
  {{K}^{ijkl}[{R}]}= {R}^{aibj}{} {^\ast}\hspace{-1pt}\!{R}
  \!\,^\ast_{\hspace{2pt}acbd}{R}^{ckdl}= {{{K}^{klij}[{R}]}\,.}
\end{eqnarray}
It has the same algebraic symmetries as $K[\chi]$. Hence, it carries
136 independent components.

For {\it vanishing torsion,} the RC-curvature $R_{ijk}{}^l$ becomes
the Riemann curvature $\widetilde{R}_{ijk}{}^l$, with the
corresponding Kummer tensor $K^{ijkl}[\widetilde{R}]$ that, for
vanishing Ricci tensor, $\widetilde{\rm R}\text{ic}_{ij}=0$,
eventually specializes to Zund's Kummer tensor
$K^{ijkl}[\widetilde{C}]$ in \cite{Zund_1976}. In each step, the
tensor fed into the ``Kummer machine'' loses some independent
components. As a consequence, the corresponding Kummer tensor picks up
additional algebraic symmetries.

\subsection{Kummer--Weyl tensor $K^{ijkl}[C]$ in Riemann-Cartan and in
  Riemann spacetime}

In a RC-space, the curvature can be decomposed under the SO(1,3) into
6 independent pieces, see \cite{PRs} for details,
\begin{equation}\label{Rdecomp}
  R_{ijk}{}^l=\sum_{I=1}^{6}\,^{(I)}\!R_{ijk}{}^l\,,\qquad 36=10\oplus
  9\oplus 1\oplus9\oplus 6\oplus 1\,.
\end{equation}
The first piece $^{(1)}\!R_{ijk}{}^l=:C_{ijk}{}^l$ is the Weyl tensor
of the RC-space with 10 independent components. It so happens that the
corresponding Weyl tensor of the Riemann space
$\widetilde{C}_{ijk}{}^l$ has 10 independent components as well.

This can be understood in the following way: the Weyl tensor
$C^{ijkl}$, as curvature tensor of a RC-space, obeys the symmetries
(\ref{SymmRC}); moreover, the irreducible pieces $^{(I)}C^{ijkl}$, for
$I=2,3,4,5,6$, have to vanish.\footnote{These vanishing pieces add up
  to $9+1+9+6+1=26$ independent pieces. Thus, for the Weyl tensor,
  $36-26=10$ independent pieces are left over.} This yields the the
algebraic symmetries of the Weyl tensor,
\begin{equation}\label{SymmRC*}
  C^{(ij)kl}=0,\quad C^{ij(kl)}=0,\quad C^{ijkl}=C^{klij},\quad C^{[ijkl]}=0,
\quad g_{kl}C^{kijl}=0\,,
\end{equation}
leaving us with $10$ independent components for the Weyl tensor
$C^{ijkl}$.

In a Riemann space, the vanishing of the torsion induces the following
symmetries for the Riemann curvature
$\widetilde{R}^{ijkl}=\widetilde{R}^{klij},\,
\widetilde{R}^{[ijkl]}=0$. As a consequence, since also
$\widetilde{\rm R}\text{ic}_{[ij]}=0$, the Weyl tensor of a Riemann
space obeys the same symmetries as the Weyl tensor of a
RC-space:
\begin{equation}\label{SymmRC**}
  \widetilde{C}^{(ij)kl}=0,\quad \widetilde{C}^{ij(kl)}=0,\quad  
  \widetilde{C}^{ijkl}= \widetilde{C}^{klij},\quad  \widetilde{C}^{[ijkl]}=0,
  \quad g_{kl} \widetilde{C}^{kijl}=0\,.
\end{equation}
Algebraically, $C^{ijkl}$ and $\widetilde{C}^{ijkl}$ share the same
symmetries. Hence, $C^{ijkl}$ and $\widetilde{C}^{ijkl}$, both have
$10$ independent components. The anti-self double duality of
$\widetilde{C}^{ijkl}$, known from GR, translates into the
corresponding symmetry of ${C}^{ijkl}$:
\begin{equation}\label{Cdual}
  C^{ijkl}=-{}^*\!C^{*\,ijkl}\,,\qquad  \widetilde{C}^{ijkl}=-{}^*\! 
  \widetilde{C}^{*\,ijkl}\,.
\end{equation}
These relations can be proven by substituting the definitions of the
dualities and by using some of the symmetries in (\ref{SymmRC*}) or
(\ref{SymmRC**}), respectively.

Let us turn to the Kummer--Weyl tensor, see the definition (\ref{KummerRC1}):
\begin{eqnarray}
  \label{KummerWeylRC}
  {{K}^{ijkl}[{C}]}= {C}^{ai bj}\hspace{1pt}
  {^*}\!{C}^*_{\hspace{2pt}acbd}\,{C}^{ckdl}\,.
\end{eqnarray}
The analogous relation is valid for vanishing torsion, that is,
for $\widetilde{C}^{\,ijkl}$. Such statements are valid for all the Kummer
formulas relating to $K[C]$. We will not mention this further.  Because
of (\ref{Cdual}), the definition simplifies appreciably:
\begin{eqnarray}
  \label{KummerRC1*}
  {{{K}^{ijkl}[{C}]}=- {C}^{aibj} {C}_{acbd} {C}^{ckdl}\,.}
\end{eqnarray}

We have again the symmetries (\ref{convSym}) and (\ref{addSymRiemann}):
\begin{equation} \label{SymKumC} {{K}^{ijkl}[{C}]}=
  {{K}^{klij}[{C}]}\,,\qquad K^{ijkl}[{C}]=K^{jilk}[{C}]\,.
\end{equation}

Since $K^{ijkl}[\widetilde{R}]$ should have more independent
components than $K^{ijkl}[C]$, we have to expect additional symmetries
for  $K^{ijkl}[C]$. Indeed, one finds that 
\begin{equation}\label{K_s3}
  K^{[ij](kl)}[C]=0\,,\qquad K^{(kl)[ij]}[C]=0\,.
\end{equation}
For the proof we expand the parentheses and the brackets:
\begin{eqnarray}
  K^{[ i j]( k l)}[C] & = & \frac{1}{2}\left(K^{[ i j] k l}
+K^{[ i j] l k} \right)
   =  \frac{1}{4}\left(K^{ i j k l}-K^{ j i k l}+K^{ i j l k}
    -K^{ j i l k}\right)\cr
  & = & \frac{1}{4}\left[ \left( K^{ i j k l} - K^{ j i l k}\right)
    -\left(K^{ j i k l}-K^{ i j l k}\right)\right]=0\,.
\end{eqnarray}
Here we made use of (\ref{SymKumC})$_2$. Accordingly, the symmetries
(\ref{K_s3}) are not independent from (\ref{SymKumC})$_2$. We collect
our results:

\noindent{\bf Proposition:} {\it The Kummer-Weyl tensors $K^{ijkl}[C]$
  and $K^{ijkl}[\widetilde{C}]$ fulfill the algebraic symmetries}
\begin{equation}
K^{ i j k l} = K^{ k l i j}\,,\quad K^{ i j k l}=K^{ j i l k}\,,
\quad  K^{[ i j]( k l)}=0\,,\quad K^{( k l)[ i j]}=0\,.
\end{equation}

\section{Canonical decomposition of ${\cal K}^{ijkl}$ 
under $GL(4,R)$}

We are looking for a decomposition of the premetric Kummer tensor
density ${\cal K}^{ijkl}$ that is irreducible under the action of the
general linear group $GL(4,R)$. For this purpose it is convenient
to turn first to a...

\subsection{General fourth rank tensor density ${\cal S}^{ijkl}$}

Due to the Schur-Weyl duality theorem, see \cite{Weyl}, the
decomposition of ${\cal S}^{ijkl}$ is related to the irreducible
finite-dimensional representations of the symmetric group $S_4$. An
interplay between the groups $S_4$ and $GL(4,R)$ determines the
decomposition of ${\cal S}^{ijkl}$.

First we have to consider the Young\footnote{Alfred Young
  (1873--1940), English mathematician and priest.} diagrams of ${\cal
  S}^{ijkl}$, see Boerner \cite{Boerner} or Hamermesh
\cite{hamermesh}. These diagrams are a graphical representations of
the $S_4$:
\begin{equation}\label{irr1}
  \Yvcentermath1 \lambda_1=\yng(4)\quad\lambda_2=\yng(3,1)\quad 
  \lambda_3=\yng(2,2)\quad\lambda_4=\yng(2,1,1)\quad\lambda_5=\yng(1,1,1,1)
\end{equation}
The Schur-Weyl duality theorem can be stated as follows: Let be given
a tensor (or a tensor density) of a rank $p$ over the space ${R}^n$.
For $p\ge 2$, the tensor space ${\cal T}^p_n$ is decomposed into a
direct sum of its subspaces,
\begin{equation}\label{irr1a}{\cal T}^p_n=
\oplus_\lambda\Big(S^\lambda_p\otimes G^\lambda_n\Big)\,.
\end{equation}
Here $S^\lambda_p$ is a representation of a direct sub-group of $S_p$
and $ G^\lambda_n$ that of $GL(n, R)$. The tensor products of pairs of
the irreducible moduli for these two groups are subspaces of ${\cal
  T}^p_n$ that are summed over all Young diagrams $\lambda_i$.

The dimension of $S^\lambda_p$ is calculated by the {\it hook-length
  formula}\footnote{See also Wikipedia
  http://en.wikipedia.org/wiki/Young$\underline{\hspace{6pt}}$tableau.}
\begin{equation}\label{irr1b} {\rm dim}\,
  S^\lambda_p=\frac{p!}{\prod_{x\in\lambda}h(x)}\,;
\end{equation}
here the {\it hook-length} $h(x)$ of a box $x$ is the number of boxes
that are in the same row to the right of it plus the number of boxes
in the same column below it plus 1. Using (\ref{irr1b}), we find for
the different Young diagrams $\lambda_i$,
\begin{equation}\label{irr1c}
{\rm dim} \,S^{\lambda_1}_4={\rm dim}\, S^{\lambda_5}_4=1,\quad
{\rm dim}\, S^{\lambda_2}_4={\rm dim}\, S^{\lambda_4}_4=3,\quad 
{\rm dim}\, S^{\lambda_3}_4=2\,.
\end{equation}

The dimension of the second factor $G^\lambda_n$ is calculated by the
{\it hook-content formula}, with the content $c(x)$ of the box
$x$---the number of its rows minus the number of its columns:
\begin{equation}\label{irr1d}
{\rm dim}\, G^\lambda_n=
{\prod_{x\in\lambda}\frac{n+c(x)}{h(x)}}\,.
\end{equation}
We obtain for the different diagrams $\lambda_i$ of (\ref{irr1}),
\begin{eqnarray}
&&{\rm dim} \,G^{\lambda_1}_4=35\,,\qquad 
{\rm dim}\, G^{\lambda_2}_4=45,\qquad
{\rm dim} \,G^{\lambda_3}_4=20,\nonumber\\ &&\label{irr1c*} \qquad \qquad 
{\rm dim} \,G^{\lambda_4}_4=15,\qquad 
{\rm dim} \,G^{\lambda_5}_4=1\,.
\end{eqnarray}

Consequently, for an arbitrary fourth rank tensor density ${\cal
  S}^{ijkl}$ the set of its 256 independent components is decomposed
into the subsets
\begin{equation}\label{coeff}
  {\bf 256=1\!\times\!{\bf 35}\,+\,3\!\times\! {\bf
      45}\,+\,2\!\times\! 
    {\bf 20    }\,+\,3\!\times\! {\bf 15}\,+\, 1\!\times\!{\bf 1}}\,.
\end{equation}

In terms of Young diagrams, the decomposition can be represented as
\begin{eqnarray}
  && \hspace{-20pt}
  \Yvcentermath1\yng(1)\otimes\yng(1)\otimes\yng(1)\otimes\yng(1)
  =\nonumber\\ &&\label{irr3}
  {\bf(1)}\yng(4)\oplus{\bf(3)} \yng(3,1) \oplus  {\bf(2)}\yng(2,2)
  \oplus{\bf(3)} \yng(2,1,1)\oplus {\bf(1)}\yng(1,1,1,1)\,.
\end{eqnarray}
The left-hand-side describes a general fourth rank tensor
(density). On the right-hand side, the {\it first} diagram represents
the completely symmetric tensor, the {\it three middle} diagrams
describe tensors that are partially symmetric and partially
antisymmetric, and the {\it last} diagram represents the completely
antisymmetric tensor. The numbers in brackets denote the dimension of
the irreducible representation, that is, the number of Young tableaux
associated with a given diagram.  

The interpretation is as follows: The 256-dimensional tensor space is
{\it uniquely} (``canonically'') decomposed into the direct sum of 5
tensor spaces with the dimensions indicated. Their explicit form can
be found, for instance, in Wade \cite{wade}:
 \begin{eqnarray}\label{Car-T1}
{^{[1]}{\cal S}}^{ijkl}&:=& {{\frac{ 1}{4!}}}\, (A+B+C+D+E)\,,\\
\label{Car-T2}{^{[2]}{\cal S}}^{ijkl}&:=& 
{{\frac 3{4!} }}\,(3A+B-D-E)\,,
\\ \label{Car-T3}{^{[3]}{\cal S}}^{ijkl}&:=& {{\frac{2}{4!}}}\,
(2A-C+2E)\,,\\
\label{Car-T4}{^{[4]}{\cal S}}^{ijkl}
&:=& {{\frac{3}{4!}}}\, (3A-B+D-E)\,,\\
\label{Car-T5}
{^{[5]}{\cal S}}^{ijkl}&:=&
 {{\frac{1}{4!}}}\, (A-B+C-D+E)\,.
\end{eqnarray}
Here, for conciseness, we suppressed the indices $i,j,k,l$ on the
right-hand sides. The numerators in the coefficients are the
dimensions given in (\ref{irr1c}). These five so-called cycles are
defined according to
\begin{eqnarray}\label{Car-ABCT}
A^{ijkl}[{\cal S}]&:=&{\cal S}^{ijkl}\,,\\
B^{ijkl}[{\cal S}]&:=&{\cal S}^{jikl}+{\cal S}^{kjil}+
{\cal S}^{ljki}+{\cal S}^{ikjl}+
{\cal S}^{ilkj}+{\cal S}^{ijlk}\,,\\
C^{ijkl}[{\cal S}]&:=&{\cal S}^{jkil}+{\cal S}^{kijl}+
{\cal S}^{jlki}+{\cal S}^{likj}+
{\cal S}^{kjli}+{\cal S}^{ljik}
\!+\!{\cal S}^{iklj}\!+\!{\cal S}^{iljk},\\
D^{ijkl}[{\cal S}]&:=&{\cal S}^{jkli}+{\cal S}^{klji}+
{\cal S}^{jlik}+{\cal S}^{lijk}+
{\cal S}^{kilj}+{\cal S}^{lkij}
\,,\\
E^{ijkl}[{\cal S}]&:=&{\cal S}^{jilk}+{\cal S}^{klij}+
{\cal S}^{lkji}\,.
\end{eqnarray}
The term $A$ corresponds to the identity operator, $B$ is the sum of
2-cycles (permutations of two indices), $C$ is used for 3-cycles, and
$D$ for 4-cycles.  The term $E$ presents the products of disjoint
2-cycles.

However, this decomposition is reducible. The tensor spaces
corresponding to the diagrams with the dimensions higher than one,
that is, $\lambda_2$, $\lambda_3$, and $\lambda_4$, see (\ref{irr3}),
are successively decomposed into the direct sum of $3$, $2$, and 3
isomorphic subspaces, respectively.  Accordingly, the 135-dimensional
tensor space, depicted in the second diagram of (\ref{irr3}), is
decomposed into a direct sum of three isomorphic subspaces of
dimension 45. Similarly, the 40-dimensional space corresponding to the
third diagram of (\ref{irr3}) is decomposed into a direct sum of two
isomorphic 20-dimensional subspaces.  Finally the 45-dimensional space
is decomposed into a direct sum of three isomorphic 15-dimensional
subspaces.

In this way, we arrive at an irreducible decomposition. Accordingly,
there are no minimal subspaces of dimensions different from those
specified above; or, in terms of tensors: there are no tensors, the
number of components of which is different from those given in
(\ref{coeff}).  Note, however, that the irreducible decomposition in
(\ref{coeff}) is not necessarily unique.  There is an infinite number
of ways to decompose the tensor spaces corresponding to $\lambda_2$,
$\lambda_3$, and $\lambda_4$ into their isomorphic subspaces.

\subsection{Specializing to the Kummer tensor density ${\cal
    K}^{ijkl}$}

The Kummer tensor density ${\cal K}^{ijkl}$ satisfies the symmetry
property ${\cal K}^{ijkl}={\cal K}^{klij}$ and can be represented as a
symmetric $16\times 16$ matrix, the dimension of which is 136.  The
question is now: Into which subspaces can this 136-dimensional space
be decomposed irreducibly under the $GL(4,R)$?

We turn to a dimensional analysis. We must have a partition of 136
into a sum of the numbers representing the dimensions of the invariant
subspaces (\ref{irr1c}). This yields the Diophantine equation
\begin{equation}\label{coeff2x}
  {\bf 136}=\a\!\times\!{\bf 35}\,\oplus\, \b\!\times\! {\bf 45}\,\oplus\,
  \g\!\times\! {\bf 20 }\,\oplus\,\d\!\times\!  {\bf 15}\, \oplus\,
  \eps\!\times\!{\bf 1}\,.
\end{equation}
The coefficients have to be less or equal to the corresponding
dimensions specified in (\ref{irr1c}). Then, $0\le\a,\eps\le 1$,
$\quad 0\le\b,\d\le 3$, and $0\le\g\le 2$. Since the coefficient
before the total symmetric part $\a=1$, the unique solution is
\begin{equation}\label{coeff2}
  \a=\b=\d=\eps=1\,,\qquad \g=2\,.
\end{equation}
Thus, we have a unique decomposition of the Kummer tensor into 5
pieces:
\begin{equation}\label{irr2}
  \Yvcentermath1\yng(1)\otimes\yng(1)\otimes\yng(1)\otimes\yng(1)
  ={\bf(1)}\yng(4)\oplus{\bf(1)} \yng(3,1) \oplus  {\bf(2)}\yng(2,2)\oplus{\bf(1)} \yng(2,1,1)\oplus {\bf(1)}\yng(1,1,1,1)\,.
\end{equation}
Using the formulas (\ref{Car-T1}) to (\ref{Car-T5}) and the symmetry
of the Kummer tensor ${\cal K}^{ijkl}={\cal K}^{klij}$, we find for
the five canonical pieces,
 \begin{eqnarray}\label{Car-K*1}
{^{[1]}{\cal K}}^{ijkl}&=&{\cal K}^{(ijkl)}\,,\\
\label{Car-K*2}{^{[2]}{\cal K}}^{ijkl}&=& 
 {
{\frac{1}{2}}}\,({\cal K}^{(i|j|k)l}-{\cal
  K}^{j(i|l|k)})  \,,
\\
\label{Car-K*3}{^{[3]}{\cal K}}^{ijkl}&=&
{
{\frac{1}{6}}}\,(2{\cal K}^{ijkl}-{\cal K}^{jkil} -{\cal
  K}^{kijl}-{\cal K}^{likj}-{\cal
  K}^{ljik}+2{\cal K}^{jilk})\,,\hspace{30pt}\\
\label{Car-K*4}{^{[4]}{\cal K}}^{ijkl}
&=&
{
{\frac 12}}\,({\cal K}^{[i|j|k]l}-{\cal K}^{j[i|l|k]})\,,\\
\label{Car-K*5}
{^{[5]}{\cal K}}^{ijkl}&=&
{\cal K}^{[ijkl]}\,.
\end{eqnarray}

Each piece has then inherited the symmetry of the Kummer tensor:
\begin{eqnarray}\label{irrSymm**}
{}^{[I]}{\cal K}^{ijkl}={}\hspace{3pt}^{[I]}{\cal K}^{klij}\,,\quad
\text{  for $I=1,2,3,4,5$}\,.
\end{eqnarray}
Let us look for the projection properties. If $^{[I]}{\cal Y}$ denotes the
five tensor operators corresponding to the decomposition in
(\ref{Car-K*1}) to (\ref{Car-K*5}), then we find ($I,J=1,\cdots ,5$):
\begin{equation}\label{Kron2_K}
  ^{[I]}\!{\cal Y} \circ{}  ^{[J]}\!{\cal Y} =  ^{[J]}\!\!{\cal Y}{\delta}^{I}_J\, \qquad \text{(no summation over $J$)}\,.
\end{equation}
The orthogonality of two operators corresponding to different diagrams
and their projection property is a generic fact \cite{Ma:2004}.

We can collect our results in the 

{\bf $\bullet$ Proposition:} {\it With the Young tableau technique,
  the Kummer tensor density ${\cal K}^{ijkl}$ can be decomposed
  uniquely into the five canonical pieces ${^{[I]}{\cal K}}^{ijkl}\:$
  of equations (\ref{Car-K*1}) to (\ref{Car-K*5}) ($I=1,...,5$). Let
  the five tensorial projection operators} $^{[I]}\!{\cal Y}$ {\it be
  given that create the canonical pieces ${}^{[I]}{\cal
    K}^{ijkl}$. Then these operators satisfy the projection equations
  (\ref{Kron2_K}).  The dimensions of the canonical pieces turn out to
  be as follows:
\begin{eqnarray} \label{canonical}{\cal K}^{ijkl} &=& {}{^{[1]}{\cal
      K}}^{ijkl} +{}{^{[2]}{\cal K}}^{ijkl}+
  {}{^{[3]}{\cal K}}^{ijkl}+{}{^{[4]}{\cal
      K}}^{ijkl}+ {}{^{[5]}{\cal
      K}}^{ijkl}\,,\nonumber\\
  136&=&\hspace{12pt}35\hspace{9pt}\oplus\hspace{13pt}
  45\hspace{8pt}\oplus \hspace{15pt}40\hspace{8pt}\oplus
  \hspace{15pt}15\hspace{8pt}\oplus\hspace{8pt}1\,.
\end{eqnarray}
Only the piece ${}{^{[3]}{\cal K}}^{ijkl}$ can be decomposed
further into a sum of two independent terms. However, this
decomposition is not uniquely defined.}

The splitting of the piece ${}{^{[3]}{\cal K}}^{ijkl}$ is the last
problem to be solved in this context.  We compute two tableaux
embraced by the term ${}{^{[3]}{\cal K}}^{ijkl}$:
 \begin{equation}\label{3a}
   \hspace{-20pt} 
   {}{^{[3a]}{\cal K}}^{ijkl}:= \Yvcentermath1\young({1}{2},{3}{4})\,
   \,{\cal K}^{ijkl}\,,\qquad {}{^{[3b]}{\cal K}}^{ijkl}:=   
   \Yvcentermath1\young({1}{3},{2}{4})\,\,{\cal K}^{ijkl}\,.
\end{equation} 
By successive permutations of the indices, we obtain 
\begin{eqnarray}\label{tab1}
  \hspace{-20pt} 
 {}{^{([3a]}{\cal K}}^{ijkl}
&=&\frac 1{12}
  \big(I+(1,2)\big)\big(I+(3,4)\big)\big(I-(1,3)\big)\big(I-(2,4)\big)
  {\cal K}^{ijkl}\nonumber\\
&=&\frac 1{12}
  \big(I+(1,2)\big)\big(I+(3,4)\big)\big(I-(1,3)\big)\big({\cal K}^{ijkl}-{\cal K}^{ilkj}\big)\nonumber\\
  &=&
\frac 1{6} \big(I+(1,2)\big)\big(I+(3,4)\big)\big( {\cal K}^{ijkl}- {\cal K}^{ilkj}\big)\nonumber\\
  &=&\frac 1{6} \big(I+(1,2)\big)\big( {\cal K}^{ijkl}- {\cal K}^{ilkj}+{\cal K}^{ijlk}- {\cal K}^{iklj} \big)\nonumber\\
  &=&\frac 13\big({\cal K}^{ij(kl)}-{\cal K}^{i(lk)j}+{\cal K}^{ji(kl)}
  -{\cal K}^{j(lk)i}\big)\,,
\end{eqnarray}
\begin{eqnarray}\label{tab2}
  \hspace{-20pt} 
  {}{^{[3b]}{\cal K}}^{ijkl}
  &=&\frac 1{12}\big(I+(1,3)\big)\big(I+(2,4)\big)\big(I-(1,2)\big) \big(I-(3,4)\big) {\cal K}^{ijkl}\nonumber\\
  &=&\frac 1{12}\big(I+(1,3)\big)\big(I+(2,4)\big)\big(I-(1,2)\big)  \big({\cal K}^{ijkl}-{\cal K}^{ijlk}\big)
  \nonumber\\
  &=&\frac 1{12}\big(I+(1,3)\big)\big(I+(2,4)\big)\big({\cal K}^{ijkl}-{\cal K}^{ijlk}-{\cal K}^{jikl}+{\cal K}^{jilk}\big)
  \nonumber\\
  &=&\frac 1{12}\big(I+(1,3)\big)
  \big({\cal K}^{ijkl}-{\cal K}^{ijlk}-{\cal K}^{jikl}+{\cal K}^{jilk}
  {\cal K}^{ilkj}-{\cal K}^{iljk}\nonumber\\ &&\hspace{70pt}-{\cal K}^{likj}+{\cal K}^{lijk}\big)
  \nonumber\\
  &=&\frac 13\big({\cal K}^{ij[kl]}+{\cal K}^{kj[il]}+{\cal
    K}^{ji[lk]}+  {\cal K}^{jk[li]}\big)\,.
 \end{eqnarray}
Observe that the sum of these terms indeed equals to $ {}{^{[3]}{\cal K}}^{ijkl}$:
\begin{equation}\label{ab-sum}
 {}{^{[3a]}{\cal K}}^{ijkl}+ {}{^{[3b]}{\cal K}}^{ijkl}={}{^{[3]}{\cal K}}^{ijkl} \,.
\end{equation} 
Moreover, both terms satisfy the symmetry of a Kummer tensor density:
\begin{equation}\label{ab-relation}
 {}{^{[3a]}{\cal K}}^{ijkl}={}{^{[3a]}{\cal K}}^{klij} \,,\qquad 
{}{^{[3b]}{\cal K}}^{ijkl}={}{^{[3b]}{\cal K}}^{klij} \,.
\end{equation} 
Accordingly, these pieces can serve as Kummer tensor densities
themselves. 

It should be remarked, however, that two operators associated with the
same Young diagram are not necessary orthogonal, see the examples in
\cite{Ma:2004}. Still, the projective property (\ref{Kron2_K}) holds
true also for the terms ${}{^{[3a]}{\cal K}}^{ijkl}$ and
${}{^{[3b]}{\cal K}}^{ijkl}$.  Indeed, the corresponding operators are
orthogonal to ${}{^{[I]}{\cal Y}}$, for $I=1,2,4,5$, since their
images are lying in ${}{^{[3]}{\cal Y}}$.  

To demonstrate the orthogonality between ${}{^{[3a]}{\cal K}}^{ijkl}$
and ${}{^{[3b]}{\cal K}}^{ijkl}$, it is enough to consider the
products of the operators specified by (\ref{tab1}) and (\ref{tab2}).
We have
\begin{equation}
{}{^{[3a]}{\cal Y}}\circ{}{^{[3b]}{\cal Y}}=\cdots\big(I-(2,4)\big)
\big(I+(2,4)\big)\cdots=0\,. 
\end{equation}
A similar relation holds for the product ${}{^{[3b]}{\cal
    Y}}\circ{}{^{[3a]}{\cal Y}}$.  The projective property ${\cal
  Y}\circ{}{\cal Y}={\cal Y}$ is a generic fact for every Young
diagram with normalized operators. Note that in \cite{Ma:2004} a
non-normalized version of the operators is used. Thus, finally, we
arrive at the...

{\bf $\bullet$ Proposition on the irreducible decomposition of the 
Kummer tensor density:}
{\it The Kummer tensor density ${\cal K}^{ijkl}[T]$ decomposes into
  six pieces according to}
\begin{eqnarray} \label{irr-canonical}{\cal K}^{ijkl} &=&
  {}{^{[1]}{\cal K}}^{ijkl} +{}{^{[2]}{\cal K}}^{ijkl}+
  {}{^{[3a]}{\cal K}}^{ijkl}+{}{^{[3b]}{\cal K}}^{ijkl}+{}{^{[4]}{\cal
      K}}^{ijkl}+ {}{^{[5]}{\cal
      K}}^{ijkl}\,,\nonumber\\
  136&=&\hspace{10pt}35\hspace{14pt}\oplus\hspace{13pt}
  45\hspace{10pt}\oplus \hspace{15pt}20\hspace{13pt}\oplus
  \hspace{13pt}20\hspace{12pt}\oplus
  \hspace{12pt}15\hspace{12pt}\oplus\hspace{12pt}1\,,\\ \nonumber
  {\cal K}^{ijkl} &=& {}{^{(1)}{\cal K}}^{ijkl} +{}{^{(2)}{\cal
      K}}^{ijkl}+ {}{^{(3)}{\cal K}}^{ijkl}+{}{^{(4)}{\cal
      K}}^{ijkl}+{}{^{(5)}{\cal K}}^{ijkl}+ {}{^{(6)}{\cal
      K}}^{ijkl}\,.
\end{eqnarray}
{\it The pieces ${}{^{(3)}{\cal K}}^{ijkl}$ and ${}{^{(4)}{\cal
      K}}^{ijkl}$ are not uniquely determined.}\medskip

The {\it new notation} in the last line with parentheses, that is,
$^{(I)}K^{ijkl}$, for $I=1,2,3,4,5,6$, has to be carefully
distinguished from $^{[J]}K^{ijkl}$, for $J=1,2,3,4,5$. We introduced
this notation because it is often more convenient to work with
successive Arabic numbers.

\subsection{Kummer tensor density ${\cal K}^{ijkl}[{\cal T}]$ 
associated with irreducible pieces of $\cal T$}

\subsubsection*{Axion piece}

The simplest example is the {axion piece} ${}^{(3)}{\cal
  T}^{ijkl}=\a\,\epsilon^{ijkl}$. Its ``diamond'' double dual
(\ref{doublediamond}) can be calculated straightforwardly:
\begin{eqnarray}
\label{spec7}
^{\diamond\,(3)}{\cal T}^\diamond_{\hspace{2pt}ijkl}=
\frac 14 \eps_{ijab}\,\a\eps^{abcd}\,\eps_{klcd}=\frac
14\a\eps_{ijab}2(\d^a_k\d^b_l-\d^a_l\d^b_k)=\a\,\eps_{ijkl}\,.
\end{eqnarray}
Thus, the corresponding Kummer tensor density takes the form
\begin{eqnarray}
\label{spec8}
{\cal K}^{ijkl}[{}^{(3)}{\cal T}]= \a^3\eps^{aibj}\eps_{acbd}\eps^{ckdl}=
2\a^3\eps^{ijkl}={}^{(6)}{\cal
  K}^{ijkl}[{}^{(3)}{\cal T}]
\,,
\end{eqnarray}
that is, it is totally antisymmetric.
\begin{center}
\includegraphics[width=20cm]{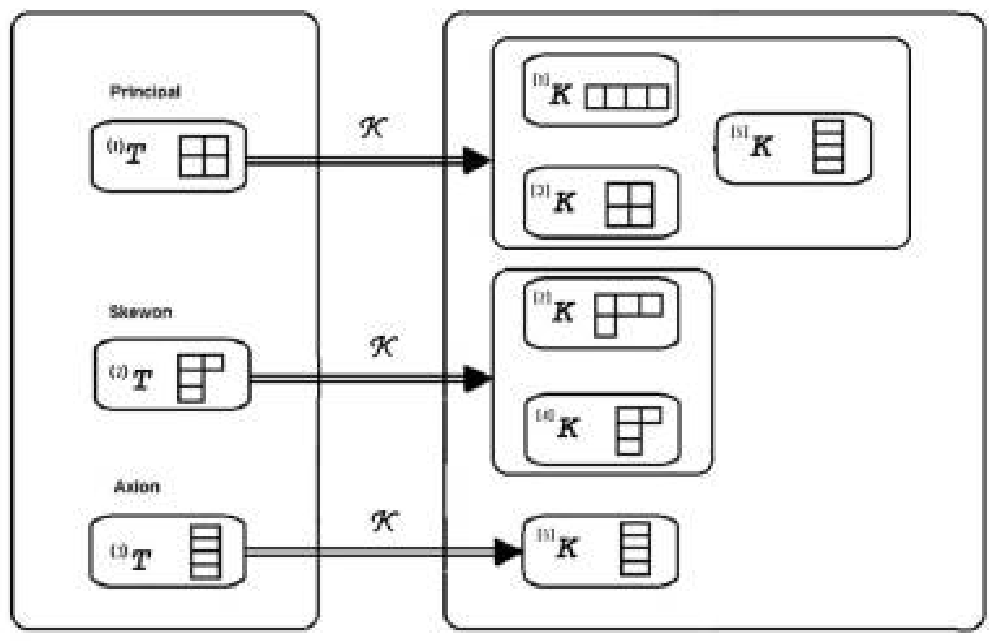}
\end{center}
\vspace{-10pt} Figure 5. {\it If a single irreducible piece
  $^{(I)}{\cal T}^{ijkl}$, for $I=1,2,\text{or }3$, of the doubly
  antisymmetric tensor density ${\cal T}^{ijkl}$ is given, we depict
  how it maps into the irreducible pieces of the Kummer tensor density
  $^{[J]}{\cal K}^{ijkl}$, for $J=1,...,5$.}

\subsubsection*{Skewon piece}

The Kummer tensor density of the {skewon piece} is a bit more
involved. First we observe that the skewon part
, by definition, satisfies the pair antisymmetry
\begin{equation}\label{spec9}   
{}^{(2)}\!{\cal T}^{ijkl}=-{}^{(2)}\!{\cal T}^{klij}\,.
 \end{equation}
The same property holds for its diamond double dual:
\begin{equation}\label{spec10}   
  ^{\diamond\,(2)}\!{\cal T}^\diamond_{\hspace{2pt}ijkl}
  =-^{\diamond\,(2)}\!  {\cal T}^\diamond_{\hspace{2pt}klij}\,.
 \end{equation}
Indeed,
\begin{eqnarray}\label{spec11}   
^{\diamond\,(2)}\!{\cal T}^\diamond_{\hspace{2pt}ijkl}&=&
\frac 14 \epsilon_{ijab}\,^{(2)}{\cal T}^{abcd}\epsilon_{klcd}=
-\frac 14 \epsilon_{klcd}\,^{(2)}{\cal T}^{cdab}\epsilon_{ijab}
=-^{\diamond\,(2)}\!{\cal T}^\diamond_{\hspace{2pt}klij}\,.
\end{eqnarray}
Eqs.(\ref{spec9}) and (\ref{spec10}) yield an additional antisymmetry of
the Kummer tensor density associated with the skewon piece:
\begin{equation}\label{spec12}   
{\cal K}^{ijkl}\big[{}^{(2)}\!{\cal T}\big] =-{\cal K}^{jilk}
\big[{}^{(2)}\!{\cal T}\big] \,.
\end{equation}
This relation is proved by applying the antisymmetries (\ref{spec9})
and (\ref{spec10}) in the definition of the Kummer tensor density and
subsequently its symmetry (\ref{paircom}):
\begin{eqnarray}\label{spec13}  
{\cal K}^{ijkl}\big[{}^{(2)}\!{\cal T}\big] &=&\quad{}^{(2)}\!{\cal T}^{aibj}
\,^{\diamond(2)}\!{\cal T}^\diamond_{\hspace{2pt}acbd}
{}^{(2)}\!{\cal T}^{ckdl}
\nonumber \\ &=&
-{}^{(2)}{\cal T}^{bjai}\,
^{\diamond(2)}{\cal T}^\diamond_{\hspace{2pt}bdac}\,^{(2)}{\cal T}^{dlck}= 
-{\cal K}^{jilk}\big[{}^{(2)}\!{\cal T}\big]\,.
\end{eqnarray}

Consequently, in the irreducible decomposition of ${\cal
  K}^{ijkl}\big[{}^{(2)}\!{\cal T}\big]$ only those pieces can appear
that obey (\ref{spec12}). This is the case with the pieces
$^{(2)}{\cal K}^{ijkl}$ and $^{(5)}{\cal K}^{ijkl}$. Thus,
\begin{equation}\label{spec14}   
{\cal K}^{ijkl}\big[{}^{(2)}\!{\cal T}\big] ={}{^{(2)}{\cal K}}^{ijkl}
\big[{}^{(2)}\!{\cal T}\big]+{}{^{(5)}{\cal K}}^{ijkl}
\big[{}^{(2)}\!{\cal T}\big]\,.
\end{equation}
In terms of Young tableaux, this decomposition is represented
by
\begin{equation}\label{spec15}  
{\cal K}^{ijkl}\big[{}^{(2)}\!{\cal T}\big] =\Yvcentermath1{\bf(1)}
\yng(3,1) \oplus  {\bf(1)} \yng(2,1,1)\,.
\end{equation}

We can also check explicitly that our conclusion ({\ref{spec14}})
   is correct.  Using the expressions (\ref{Car-K*1}--\ref{Car-K*5}),
   we calculate
\begin{eqnarray*}
{^{[1]}{\cal K}}^{ijkl}&=&{\cal K}^{(ijkl)}=0\,,\\
{^{[2]}{\cal K}}^{ijkl}&=& 
 {
{\frac{1}{4}}}({\cal K}^{ijkl}+{\cal K}^{kjil}-{\cal
  K}^{jilk}-{\cal K}^{jkli})  ={\cal K}^{(i|j|k)l}\ne 0\,,\\
{^{[3]}{\cal K}}^{ijkl}
&=&{\frac{1}{6}}(2{\cal K}^{ijkl}-{\cal K}^{jkil} -{\cal K}^{kijl} 
-{\cal K}^{likj}-{\cal K}^{ljik} +2{\cal K}^{jilk}) \nonumber\\
&=&{\frac{1}{6}}(2({\cal K}^{ijkl}+{\cal K}^{jilk})-({\cal K}^{jkil}
+{\cal K}^{likj})-({\cal K}^{kijl}+{\cal K}^{ljik})=0\,,\\
{^{[4]}{\cal K}}^{ijkl}&=&{\frac 14}({\cal K}^{ijkl}-{\cal K}^{kjil}
-{\cal K}^{jilk}+{\cal K}^{jkli})
={\cal K}^{[i|j|k]l}\ne 0\,,\\
{^{[5]}{\cal K}}^{ijkl}&=&{\cal K}^{[ijkl]}=0\,.
\end{eqnarray*}

\subsubsection*{Principal piece}

It is characterized by two symmetry relations
\begin{equation}\label{spec16}  
{}^{(1)}\!{\cal T}^{ijkl}={}^{(1)}\!{\cal T}^{klij}\,,
\end{equation}
and
\begin{equation}\label{spec17}  
{}^{(1)}\!{\cal T}^{[ijkl]}=0\,.
\end{equation}
Again, its diamond double dual reflects the symmetry (\ref{spec16}),
\begin{equation}\label{spec18}   
  ^{\diamond(1)}{\cal T}^\diamond_{\hspace{2pt}ijkl}
  ={}^{\diamond(1)}{\cal T}^\diamond_{\hspace{2pt}klij}\,,
 \end{equation}
as can be seen by the definition of the double dual and by using (\ref{spec16}):
\begin{eqnarray}\label{spec19}   
^{\diamond(1)}\!{\cal T}^\diamond_{\hspace{2pt}ijkl}&=&
\frac 14 \epsilon_{ijab}{}^{(1)}{\cal T}^{abcd}\epsilon_{klcd}=
\frac 14 \epsilon_{klcd}{}^{(1)}{\cal T}^{cdab}\epsilon_{ijab}
={}^{\diamond(1)}\!{\cal T}^\diamond_{\hspace{2pt}klij}\,.
\end{eqnarray}

Similar as in the skewon case, the symmetries (\ref{spec16}) and
(\ref{spec18}) yield the relation
\begin{equation}\label{spec20}   
{\cal K}^{ijkl}\big[{}^{(1)}\!{\cal T}\big] ={\cal K}^{jilk}\big[{}^{(1)}\!{\cal T}\big] \,.
 \end{equation}
 Only the irreducible pieces ${}{^{(1)}{\cal K}}^{ijkl}$,
 ${}{^{(3)}{\cal K}}^{ijkl}$, ${}{^{(4)}{\cal K}}^{ijkl}$, and
 ${}{^{(6)}{\cal K}}^{ijkl}$ carry this symmetry. Consequently, ${\cal
   K}^{ijkl}\big[{}^{(1)}\!{\cal T}\big]$ can include only these
 pieces. Accordingly, we are left with the decomposition
\begin{equation}\label{spec21}   
{\cal K}^{ijkl}\big[{}^{(1)}\!{\cal T}\big] ={}{^{(1)}{\cal
    K}}^{ijkl}\big[{}^{(1)}\!{\cal T}\big]\!
+{}{^{(3)}{\cal K}}^{ijkl}\big[{}^{(1)}\!{\cal T}\big]\!
+{}{^{(4)}{\cal K}}^{ijkl}\big[{}^{(1)}\!{\cal T}\big]\!
+{}{^{(6)}{\cal K}}^{ijkl}\big[{}^{(1)}\!{\cal T}\big]\,.
\end{equation}
In the notation of Young tableaux, this decomposition is represented
by
\begin{equation}\label{spec22} {\cal K}^{ijkl}\big[{}^{(1)}\!{\cal
    T}\big] =\Yvcentermath1{\bf(1)}\yng(4) \oplus {\bf(2)}\yng(2,2)
  \oplus {\bf(1)}\yng(1,1,1,1)\,.
 \end{equation}

 \section{Presence of a metric and 
   decomposition of $K^{\a\b\g\d}$ under $SO(1,3)$}

As soon as a metric is available, we can, according to (\ref{dens}),
pass over from the Kummer tensor density ${\cal K}^{ijkl}$ to the
Kummer tensor $K^{ijkl}$. Its six irreducible $GL(4,R)$ pieces
$^{(I)}{K}^{ijkl}$ can now be decomposed still finer yielding pieces
that are invariant under the Lorentz group $SO(1,3)$.

Having now locally the $SO(1,3)$ available, it is useful for later
applications to introduce a {\it local orthonormal frame} ${\bf
  e}_\a=e^{i}{}_{\a}\,{\partial}_{i}$, with ${\bf e}_\a\cdot {\bf
  e}_\b =g_{\a\b}\stareq{\rm diag}(-1,1,1,1)$. Besides the Latin
(holonomic) coordinate indices $i,j,k,\cdots$, we have now Greek
(anholonomic) frame indices $\a,\b,\g,\cdots$. The metric transforms
according to $g_{\a\b}=e^{i}{}_{\a}e^{j}{}_{\b}\,g_{ij}$. The dual
coframe $\boldsymbol{\vt}^\a=e_i{}^\a dx^i$ is then also orthonormal,
with $e_k{}^\a\,e^k{}_\b=\d^\a_\b$ and $e_i{}^\g\,e^j{}_\g=\d_i^j$.

We will refer the Kummer tensor to the orthonormal frame,
\begin{equation}\label{anholKummer}
K^{\a\b\g\d}=e_i{}^\a e_j{}^\b e_k{}^\g e_l{}^\d K^{ijkl}\,.
\end{equation}

In the subsequent sections, ${T}^{\a\b\g\d}$ in $K^{\a\b\g\d}[T]$ will
become the Riemann--Cartan curvature tensor $R^{\a\b\g\d}$, which is
originally defined as a {\it 2-form} according to $R_\g{}^\d=\frac
12R_{\a\b\g}{}^\d \vt^\a\wedge \vt^\b$. Then one can apply the Hodge
dual $^\star$ to it, $^\star\!  R_{\g\d}$, and the Lie dual
$^{\color{red}(*)}$ to the Lie-algebra indices $\g,\d$, namely
$R^{{\color{red}(*)}\a\b}:=\frac 12 R_{\g\d}\,\eta^{\a\b\g\d}$, for
details see \cite{PRs}. In the present section, just read $^*$ (the
ordinary dual of tensor calculus) for both stars, for the Hodge star
$^\star$ as well as for the Lie star $^{\color{red}(*)}$.

\subsection{First and second contractions of the Kummer tensor
  $K^{\a\b\g\d}[T]$}

As a 4th rank tensor, the Kummer tensor ${K}^{\a\b\g\d}[T]$ has {\it
six} possible contractions:
\begin{eqnarray} \label{SIX} {}_{{\color{red}(1)}}\Phi^{\g\d} &\!\!:=\!\!&
  g_{\a\b}{K}^{\a\b\g\d} ,\quad {}_{{\color{red}(2)}}\Phi^{\a\b}:=
  g_{\g\d}{K}^{\a\b\g\d} ,\quad
  {}_{{\color{red}(3)}}\Phi^{\b\d}:= g_{\a\g}{K}^{\a\b\g\d}  ,\nonumber\\
  {}_{{\color{red}(4)}}\Phi^{\a\d} &\!\!:=\!\!& g_{\b\g}{K}^{\a\b\g\d} ,\quad
  {}_{{\color{red}(5)}}\Phi^{\b\g}:= g_{\a\d}{K}^{\a\b\g\d} ,\quad
  {}_{{\color{red}(6)}}\Phi^{\a\g}:= g_{\b\d}{K}^{\a\b\g\d} .
\end{eqnarray}
In other words, we have $_{{\color{red}(A)}}\Phi^{\a\b}$, for $A=1,\cdots,6$; here
we used for the contractions the mapping $\{1,2\}\rightarrow 1$,
$\{3,4\}\rightarrow 2$, $\{1,3\}\rightarrow 3$, $\{2,3\}\rightarrow
4$, $\{1,4\}\rightarrow 5$, and $\{2,4\}\rightarrow 6$, which can be
read off directly from (\ref{SIX}). The symmetry (\ref{paircom}) is
also valid in anholonomic indices,
\begin{equation}\label{add1}
K^{\a\b\g\d}=K^{\g\d\b\a}\,,
\end{equation}
as can be seen from (\ref{anholKummer}).  Performing the contractions
(\ref{SIX}) on both sides of (\ref{add1}), we find
\begin{eqnarray} \label{add2} {}_{{\color{red}(1)}}\Phi^{\g\d}
  \!=\!{}_{{\color{red}(2)}}\Phi^{\g\d},\;
  {}_{{\color{red}(3)}}\Phi^{\b\d} \!
=\!{}_{{\color{red}(3)}}\Phi^{\d\b},\;
{}_{{\color{red}(4)}}\Phi^{\a\d} \!=\!{}_{{\color{red}(5)}}\Phi^{\d\a},\;
{}_{{\color{red}(6)}}\Phi^{\a\g} \!=\!{}_{{\color{red}(6)}}\Phi^{\g\a}.
\end{eqnarray}
Consequently there are six independent first contractions of the full
Kummer tensor, namely
\begin{equation}\label{add6}
  {}_{{\color{red}(1)}}\Phi^{(\a\b)},\;{}_{{\color{red}(1)}}\Phi^{[\a\b]},
\;{}_{{\color{red}(3)}}\Phi^{(\a\b)},\;
  {}_{{\color{red}(4)}}\Phi^{(\a\b)},\;{}_{{\color{red}(4)}}\Phi^{[\a\b]},
\;{}_{{\color{red}(6)}}\Phi^{(\a\b)},
\end{equation}
which could be used to define a basis for the first contractions of
the Kummer tensor.

Because of the pair symmetry (\ref{add1}), the second contractions
do not turn out to be independent:
\begin{align}\label{zusatz4}
\begin{aligned}
    {}_{{\color{red}(1)}}\Phi^{\a\b}&=K^{\mu}{}_{\mu}{}^{\a\b} \quad\Longrightarrow &
    {}_{{\color{red}(1)}}
    \Phi^{\rho}{}_{\rho}&=K^{\mu}{}_{\mu}{}^{\lambda}{}_{\lambda}=:M\,,\\
    {}_{{\color{red}(2)}}\Phi^{\a\b}&=K^{\a\b\mu}{}_{\mu}  \quad \Longrightarrow& 
    {}_{{\color{red}(2)}}
    \Phi^{\rho}{}_{\rho}&=K^{\mu}{}_{\mu}{}^{\lambda}{}_{\lambda}=M\,, \\
    {}_{{\color{red}(3)}}\Phi^{\a\b}&=K^{\mu\a}{}_{\mu}{}^{\b}   \quad\Longrightarrow&  
    {}_{{\color{red}(3)}}\Phi^{\rho}{}_{\rho}&=K^{\mu\lambda}{}_{\mu\lambda} =: L\,, \\
    {}_{{\color{red}(4)}}\Phi^{\a\b}&=K^{\a\mu}{}_{\mu}{}^{\b}  \quad \Longrightarrow&  
    {}_{{\color{red}(4)}}\Phi^{\rho}{}_{\rho}&=K^{\lambda\mu}{}_{\mu\lambda} =: K\,, \\
    {}_{{\color{red}(5)}}\Phi^{\a\b}&=K^{\mu\a\b}{}_{\mu}  \quad \Longrightarrow&
    {}_{{\color{red}(5)}}\Phi^{\rho}{}_{\rho}&=K^{\mu\lambda}{}_{\lambda\mu} = K\,, \\
    {}_{{\color{red}(6)}}\Phi^{\a\b}&=K^{\a\mu\b}{}_{\mu}  \quad \Longrightarrow&  
    {}_{{\color{red}(6)}}\Phi^{\rho}{}_{\rho}&=K^{\mu\lambda}{}_{\mu\lambda} = L\,.
\end{aligned}
\end{align}
Thus, the Kummer tensor entails the three independent scalars $\{K\,,
L\,, M\}$. If the ``Kummer machinery'' is fed with a more symmetric
tensor, the Weyl tensor $C_{\a\b\gamma}{}^{\delta}$, for instance, we
find further algebraic relations between these scalars. In the case of
$T=C$, only one independent scalar, say $K$, is left over.

The analogous contractions can be made for each irreducible piece
$\,^{(I)}{K}^{\a\b\g\d}[T]$ of the Kummer tensor. We find (see Fig.6),
\begin{equation}\label{zusatz3}
  ^{(I)}_{{\color{red}(A)}}{\Phi}^{\a\b}\,, \quad I=1,\cdots,6\,,
  \quad
  {\color{red}A}=1,\cdots,6\,.
\end{equation}

\begin{center}
\includegraphics[width=10truecm]
{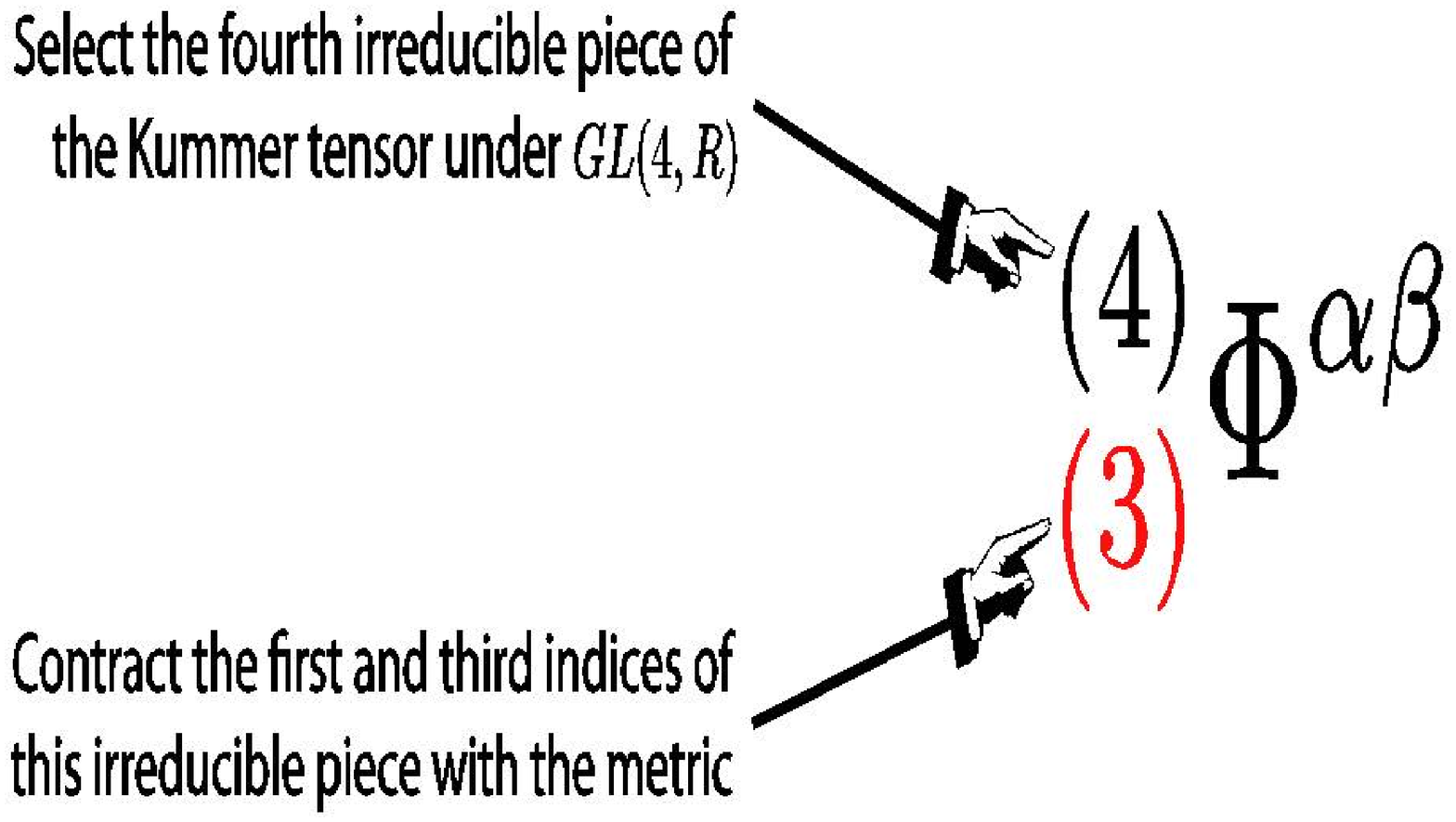}
\end{center}
\medskip

Figure 6. {\it Our use of the different types of indices: The 4th
  irreducible piece of the Kummer tensor density can be found in
  Eq.(\ref{irr-canonical}) together with Eq.(\ref{tab2}). The number 3
  denotes, according to Eq.(\ref{SIX}), a contraction over its first
  and third index.}\medskip\bigskip

Since $^{(6)}K^{\a\b\g\d}$ is totally antisymmetric, its traces vanish
and we have
\begin{equation}\label{zero}
  ^{(6)}_{{\color{red}(A)}}\!{\Phi}^{\a\b}\equiv 0\,,\quad\text{for}\quad
  {\color{red}A}=1,\cdots,6\,.
\end{equation}
Thus, in the future, we need to number the index $I$ only up to $5$.
Since each irreducible piece of $K$ has six traces and the traces of
$^{(6)}K$ are identically zero, the first contractions yield
altogether $5\times 6=30$ potentially nonvanishing traces. Because of
the existing symmetries of the $^{(I)}K^{\a\b\g\d}$, not all of these
contractions will be independent. With the help of the computer
algebra package Reduce-Excalc, see
\cite{Johannes,Hearn,Socorro,reduce93}, we were led to the results
collected in Table 1.\bigskip

\centerline{\bf Table 1. First
  and second contractions of {\em K\hspace{-1pt}ummer}}

\begin{equation*}\label{tab_Phi_s}
\begin{tabular}{|c||c|}\hline
  & ${K}^{\a\b\gamma\delta}[T] = T^{\mu\a\lambda\b}\,
  ^{\star}T^{\rstar}{}_{\hspace{-3pt}\mu\rho\lambda\sigma}T^{\rho\gamma\sigma\delta}$
  \cr\hline\hline
  $^{(1)}{K}^{\a\b\gamma\delta}$ &
  $^{(1)}_{{\color{red}(1)}}{\Phi}^{\a\b}=\,
  ^{(1)}_{{\color{red}(2)}}{\Phi}^{\a\b}=\,
  ^{(1)}_{{\color{red}(3)}}{\Phi}^{\a\b}=\,
  ^{(1)}_{{\color{red}(4)}}{\Phi}^{\a\b}=\, ^{(1)}_{{\color{red}(5)}}{\Phi}^{\a\b}=\,
  ^{(1)}_{{\color{red}(6)}}{\Phi}^{\a\b}\neq 0$ \cr\hline
  & ${\phi}_{(1)}:=\, ^{(1)}_{{\color{red}(A)}}{\Phi}^{\mu}{}_{\mu}\, \quad
  (A=1,\cdots,6)$\cr\hline
  $^{(2)}{K}^{\a\b\gamma\delta}$ &
  $^{(2)}_{{\color{red}(2)}}{\Phi}^{\a\b}=\,
  ^{(2)}_{{\color{red}(1)}}{\Phi}^{\a\b}$\,, $^{(2)}_{{\color{red}(3)}}{\Phi}^{\a\b}=-
  \, ^{(2)}_{{\color{red}(6)}}{\Phi}^{\a\b}$ \,, $^{(2)}_{{\color{red}(4)}}{\Phi}^{\a\b}=\,
  ^{(2)}_{{\color{red}(1)}}{\Phi}^{\a\b}$\,,\cr\hline
  & $^{(2)}_{{\color{red}(5)}}{\Phi}^{\a\b} = -\, ^{(2)}_{{\color{red}(1)}}{\Phi}^{\a\b}$
  \cr\hline
  & ${\phi}_{(2)}:=\, ^{(2)}_{{\color{red}(A)}}{\Phi}^{\mu}{}_{\mu}=0\,
  \quad   (A=1,\cdots , 6)$
  \cr\hline
  $^{(3)}{K}^{\a\b\gamma\delta}$ & $^{(3)}_{{\color{red}(2)}}{\Phi}^{\a\b}=\,
  ^{(3)}_{{\color{red}(4)}}{\Phi}^{\a\b}=\,^{(3)}_{{\color{red}(5)}}{\Phi}^{\a\b}=\,
  ^{(3)}_{{\color{red}(1)}}{\Phi}^{\a\b}$\,, \cr\hline
  & $^{(3)}_{{\color{red}(3)}}{\Phi}^{\a\b}=\,
  ^{(3)}_{{\color{red}(6)}}{\Phi}^{\a\b} = -2\,^{(3)}_{{\color{red}(1)}}{\Phi}^{\a\b}$
  \cr\hline
  & ${\phi}_{(3)}:=\, ^{(3)}_{{\color{red}(1)}}{\Phi}^{\mu}{}_{\mu}$\cr\hline
  $^{(4)}{K}^{\a\b\gamma\delta}$ & $^{(4)}_{{\color{red}(1)}}{\Phi}^{\a\b}=\,
  ^{(4)}_{{\color{red}(2)}}{\Phi}^{\a\b}\,, \,
  ^{(4)}_{{\color{red}(4)}}{\Phi}^{\a\b}=\, ^{(4)}_{{\color{red}(5)}}{\Phi}^{\a\b} =
  \, - \, ^{(4)}_{{\color{red}(1)}}{\Phi}^{\a\b}\,,$\cr\hline
  & $^{(4)}_{{\color{red}(3)}}{\Phi}^{\a\b}=\,
  ^{(4)}_{{\color{red}(6)}}{\Phi}^{\a\b}   = 0$ \cr\hline
  & ${\phi}_{(4)}:=\, ^{(4)}_{{\color{red}(1)}}{\Phi}^{\mu}{}_{\mu}$\cr\hline
  $^{(5)}{K}^{\a\b\gamma\delta}$ & $^{(5)}_{{\color{red}(2)}}{\Phi}^{\a\b}=\,
  ^{(5)}_{{\color{red}(1)}}{\Phi}^{\a\b}$\,, $^{(5)}_{{\color{red}(5)}}{\Phi}^{\a\b}=\,
  ^{(5)}_{{\color{red}(1)}}{\Phi}^{\a\b}$\,, $^{(5)}_{{\color{red}(4)}}{\Phi}^{\a\b}=-\,
  ^{(5)}_{{\color{red}(1)}}{\Phi}^{\a\b}\,,$
  \cr\hline
  & $^{(5)}_{{\color{red}(3)}}{\Phi}^{\a\b}= 0$\,, \quad
  $^{(5)}_{{\color{red}(6)}}{\Phi}^{\a\b}= 0$
  \cr\hline
  & ${\phi}_{(5)}:=\, ^{(5)}_{{\color{red}(A)}}{\Phi}^{\mu}{}_{\mu}=0\,
  \quad   (A=1,\cdots , 6)$
  \cr\hline\hline
\end{tabular}
\end{equation*}

In Table 1, the first contractions of $^{(2)}K^{\a\b\g\d}$ can be
written slightly more compactly as
\begin{equation}\label{2Kcompact}
  ^{(2)}_{{\color{red}(1)}}{\Phi}^{\a\b}=^{(2)}_{{\color{red}(2)}}{\Phi}^{\a\b}=
  ^{(2)}_{{\color{red}(4)}}{\Phi}^{\a\b}=-^{(2)}_{{\color{red}(5)}}{\Phi}^{\a\b}\,,\quad
  ^{(2)}_{{\color{red}(3)}}{\Phi}^{\a\b}=- \,
  ^{(2)}_{{\color{red}(6)}}{\Phi}^{\a\b}\,.
\end{equation}
Inspection of Table~1 
shows that there are {six} independent {first contractions} of the
Kummer tensor, namely one from each of $^{(1)}K$, $^{(3)}K$,
$^{(4)}K$, and $^{(5)}K$, two from $^{(2)}{K}$, and none from
$^{(6)}{K}$. As a basis for the first contractions
$^{(I)}_{\color{red}(A)}{\Phi}^{\a\b}$, we use the set,
\begin{equation}\label{set_1}
\left\{ ^{(1)}_{\color{red}(1)}{\Phi}^{\a\b}\,, \, ^{(2)}_{\color{red}(1)}{\Phi}^{\a\b}\,, \,
^{(2)}_{\color{red}(3)}{\Phi}^{\a\b}\,, \, ^{(3)}_{\color{red}(1)}{\Phi}^{\a\b}\,, \,
^{(4)}_{\color{red}(1)}{\Phi}^{\a\b}\,, \, ^{(5)}_{\color{red}(1)}{\Phi}^{\a\b}
\right\}\,.
\end{equation}
This set corresponds to one realization. Then, with Reduce-Excalc, we
find, the following

{\bf $\bullet$ Proposition on first contractions of 
{\em K\hspace{-1pt}ummer:}}
{\it Define a tensor ${\Lambda}^{\a\b}$ as a superposition of the
  elements of the set (\ref{set_1}) according to
  \begin{equation}\label{zusatz2} {\Lambda}^{\a\b} := {\xi}_{1}\,
    ^{(1)}_{\color{red}(1)}{\Phi}^{\a\b} + {\xi}_{2}\,
    ^{(2)}_{\color{red}(1)}{\Phi}^{\a\b} + {\xi}_{3}\,
    ^{(2)}_{\color{red}(3)}{\Phi}^{\a\b} + {\xi}_{4}\,
    ^{(3)}_{\color{red}(1)}{\Phi}^{\a\b} + {\xi}_{5}\,
    ^{(4)}_{\color{red}(1)}{\Phi}^{\a\b} + {\xi}_{6}\,
    ^{(5)}_{\color{red}(1)}{\Phi}^{\a\b}\,,
\end{equation}
with arbitrary constants ${\xi}_{I}$, for $I=1,\cdots ,6$. Then
${\Lambda}^{\a\b}=0$ is equivalent to
\begin{equation}\label{zusatz1} {\xi}_{I}=0\,, \qquad \text{for}
\; I=1,\cdots ,6\,.
\end{equation}
Thus, the elements of set (\ref{set_1}) are algebraically
independent. Reversely, every 2nd rank tensor, constructed by
contraction of the Kummer tensor, can be represented as the linear
combination (\ref{zusatz2}).  Thus, the elements of set
(\ref{set_1}) span the vector space of the 2nd rank tensors.  }

Let us now turn in Table 1 to the {second contractions} of Kummer,
the traces. Because the traces
$^{(2)}_{{\color{red}(A)}}{\Phi}^{\mu}{}_{\mu}$ and
$^{(5)}_{{\color{red}(A)}}{\Phi}^{\mu}{}_{\mu}$ vanish identically for
all $A=1,\cdots ,6$, we will have only {three} independent
scalars. As the corresponding set we choose
\begin{equation}\label{THREE}
\{{\phi}_{(1)}\,,
{\phi}_{(3)}\,, {\phi}_{(4)}\}.
\end{equation}
These 3 scalars are independent linear combinations of the scalars
$K,L,M$ defined in (\ref{zusatz4}).

{\bf $\bullet$ Proposition on second contractions of
 {\em K\hspace{-1pt}ummer:}}
{\it The Kummer tensor $K^{\a\b\g\d}[T]$ has three algebraically
  independent second contraction, such as $\{{\phi}_{(1)}\,,
  {\phi}_{(3)}\,, {\phi}_{(4)}\}$.}\medskip

For later use, we collect the antisymmetric pieces of the first
contractions of {\em K\hspace{-1pt}ummer} in Table 2.\bigskip

\centerline{\bf Table 2. Antisymmetric parts of first contractions of
  {\em K\hspace{-1pt}ummer}}\vspace{-10pt}
\begin{equation*}\label{tab_Phi_a}
\begin{tabular}{|c||c|}\hline
  & ${K}^{\a\b\gamma\delta}[T] = T^{\mu\a\lambda\b}\,
  ^{\star}T^{\rstar}{}_{\hspace{-3pt}\mu\rho\lambda\sigma}T^{\rho\gamma\sigma\delta}$
  \cr\hline\hline
  $^{(1)}{K}^{\a\b\gamma\delta}$ & $^{(1)}_{{\color{red}(A)}}{\Phi}^{[\a\b]}=0 \quad
  (A=1,\cdots ,6)$
  \cr\hline
  $^{(2)}{K}^{\a\b\gamma\delta}$ & $^{(2)}_{{\color{red}(1)}}{\Phi}^{[\a\b]}\neq 0\,, \,
  ^{(2)}_{{\color{red}(2)}}{\Phi}^{[\a\b]}\neq 0\,, \,
  ^{(2)}_{{\color{red}(4)}}
  {\Phi}^{[\a\b]}\neq 0\,,
  \, ^{(2)}_{{\color{red}(5)}}{\Phi}^{[\a\b]}\neq 0 \,,$
  \cr\hline
  & $^{(2)}_{{\color{red}(1)}}{\Phi}^{[\a\b]}=\, ^{(2)}_{{\color{red}(2)}}{\Phi}^{[\a\b]}=\,
  ^{(2)}_{{\color{red}(4)}}{\Phi}^{[\a\b]}=-\,^{(2)}_{{\color{red}(5)}}{\Phi}^{[\a\b]}\,,$
  \cr\hline
  & $^{(2)}_{{\color{red}(3)}}{\Phi}^{[\a\b]}= \,^{(2)}_{{\color{red}(6)}}{\Phi}^{[\a\b]}=0$
  \cr\hline
  $^{(3)}{K}^{\a\b\gamma\delta}$ & $^{(3)}_{{\color{red}(A)}}\Phi^{[\a\b]}=0 \quad
  (A=1,\cdots ,6)$
  \cr\hline
  $^{(4)}{K}^{\a\b\gamma\delta}$ & $^{(4)}_{{\color{red}(A)}}\Phi^{[\a\b]}=0 \quad
  (A=1,\cdots ,6)$
  \cr\hline
  $^{(5)}{K}^{\a\b\gamma\delta}$ & $^{(5)}_{{\color{red}(1)}}{\Phi}^{[\a\b]}\neq 0\,, \,
  ^{(5)}_{{\color{red}(2)}}{\Phi}^{[\a\b]}\neq 0\,, \,
  ^{(5)}_{{\color{red}(4)}}
  {\Phi}^{[\a\b]}\neq 0\,,
  \, ^{(5)}_{{\color{red}(5)}}{\Phi}^{[\a\b]}\neq 0 \,,$
  \cr\hline
  & $^{(5)}_{{\color{red}(1)}}{\Phi}^{[\a\b]}=\, ^{(5)}_{{\color{red}(2)}}{\Phi}^{[\a\b]}=
  -\, ^{(5)}_{{\color{red}(4)}}{\Phi}^{[\a\b]}= \, ^{(5)}_{{\color{red}(5)}}{\Phi}^{[\a\b]}\,,$
  \cr\hline
  & $^{(5)}_{{\color{red}(3)}}{\Phi}^{[\a\b]}= \,^{(5)}_{{\color{red}(6)}}{\Phi}^{[\a\b]}=0$\,
  \cr\hline
\end{tabular}
\end{equation*}

\subsection{Completely tracefree parts 
of {\em K\hspace{-1pt}ummer} $K^{\a\b\g\d}[T]$}

In order to form $SO(1,3)$-invariant tensors from the
$GL(4,R)$-invariant tensors $^{(I)}{K}^{\a\b\g\d}[T]$, we have to
subtract out all traces. Thus, we need the maximal
number of the independent tracefree symmetric
contractions $^{(I)}_{{\color{red}(A)}}{\widehat{\Phi}}^{(\a\b)}$, their
antisymmetric counterparts $^{(I)}_{{\color{red}(A)}}{\Phi}^{[\a\b]}$, and the
traces ${\phi}_{(I)}$. Here, tracefree objects
will be denoted by a hat,
\begin{equation}\label{Ric_trf}
^{(I)}_{{\color{red}(A)}}{\widehat{\Phi}}^{\a\b}:={}
^{(I)}_{{\color{red}(A)}}\!{\Phi}^{\a\b}-\frac{1}{4}\,
^{(I)}_{{\color{red}(A)}}\!{\Phi}^{\mu}{}_{\mu}\,g^{\a\b} \quad
(\text{for  }I\;\text{and}\;A =1,\cdots,5 )\,.
\end{equation}
Furthermore, we have to take into consideration all symmetrized or
antisymmetrized third rank tensors that can be built from
$^{(I)}{\widehat{K}}^{\a\b\g\d}$.

For the completely tracefree parts of $^{(I)}{K}^{\a\b\gamma\delta}$,
we find
\begin{eqnarray}\label{K1_SO3_1}
 ^{(1)}{\widehat {K}}^{\a\b\gamma\delta} &=&\hspace{-6pt}   ^{(1)}{K}^{\a\b\gamma\delta} 
  -\,\frac{3}{4}\, {^{(1)}_{\color{red}(1)}{\widehat {\Phi}}^{(\a\b}g^{\gamma\delta)}}
  - \frac{1}{8}\, {\phi}_{(1)}\,g^{(\a\b}g^{\gamma\delta)} \,,\\
\label{K2_SO3_1}
  ^{(2)}{\widehat {K}}^{\a\b\gamma\delta} & = &\hspace{-6pt}  ^{(2)}{K}^{\a\b\gamma\delta}
  +\frac{1}{4}\left( ^{(2)}_{\color{red}(3)}{\widehat {\Phi}}^{\a\gamma}g^{\b\delta}
    - \, ^{(2)}_{\color{red}(3)}{\widehat
      {\Phi}}^{\b\delta}g^{\a\gamma} \right)\cr 
 & & - \frac{1}{3}\left(
    ^{(2)}_{\color{red}(1)}{\widehat {\Phi}}^{\a(\b|}g^{\gamma|\delta)}+\,
    ^{(2)}_{\color{red}(1)}{\widehat {\Phi}}^{\gamma(\delta|}g^{\a|\b)}
  \right)\,,\\
\label{K3_SO3_1}
  ^{(3)}{\widehat {K}}^{\a\b\gamma\delta} & = &\hspace{-6pt}   ^{(3)}{K}^{\a\b\gamma\delta}
  +\, ^{(3)}_{\color{red}(1)}{\widehat {\Phi}}^{\a\g}g^{\b\d}+\,
  ^{(3)}_{\color{red}(1)}{\widehat {\Phi}}^{\b\d}g^{\a\g}\cr
  & & -\, ^{(3)}_{\color{red}(1)}{\widehat {\Phi}}^{\a(\b|}g^{\g|\d)}
  - \, ^{(3)}_{\color{red}(1)}{\widehat {\Phi}}^{\g(\b|}g^{\a|\d)}
   + \frac{1}{6}{\phi}_{(3)}g^{\a[\b|}g^{\g|\d]}\,,\\
\label{K4_SO3_1}
^{(4)}{\widehat {K}}^{\a\b\gamma\delta}& = &\hspace{-6pt} ^{(4)}{K}^{\a\b\gamma\delta}-\,
^{(4)}_{\color{red}(1)}{\widehat {\Phi}}^{\a[\b|}g^{\g|\d]}-\,
^{(4)}_{\color{red}(1)}{\widehat {\Phi}}^{\g[\d|}g^{\a|\b]}
- \frac{1}{6}{\phi}_{(4)}g^{\a[\b|}g^{\g|\d]}\!,\\
\label{K5_SO3_1}
  ^{(5)}{\widehat {K}}^{\a\b\gamma\delta} &=&\hspace{-6pt}   ^{(5)}{K}^{\a\b\gamma\delta}-
  {^{(5)}_{\color{red}(1)}{\widehat {\Phi}}^{\a[\b|}g^{\gamma|\delta]} } - \,
  {^{(5)}_{\color{red}(1)}{\widehat {\Phi}}^{\gamma[\delta|}g^{\a|\b]} }\,,\\
\label{K6_SO3_1}
  ^{(6)}{\widehat {K}}^{\a\b\gamma\delta} &=&\hspace{-6pt}
  ^{(6)}{K}^{\a\b\gamma\delta} = {K}^{[\a\b\gamma\delta]}\,.
\end{eqnarray}

The first piece in (\ref{K1_SO3_1}), which, in the electromagnetic case,
corresponds to the Tamm--Rubilar tensor, is already irreducibly
decomposed under the $SO(1,3)$:

{\bf $\bullet$ Proposition on the irreducible decomposition of the
  totally symmetric part of {\em K\hspace{-1pt}ummer:}} {\it
  $K^{(\a\b\g\d)}[T]={}^{(1)}{\widehat{K}}^{\a\b\gamma\delta}[T]$---and
  with it in the electromagnetic case the Tamm-Rubilar tensor density
  ${\cal G}^{\a\b\g\d}[\chi]$---decompose irreducibly as follows:}
\begin{eqnarray}\label{K1_SO3_1*}
  ^{(1)}{ {K}}^{\a\b\gamma\delta} &=&\nonumber \,
  ^{(1)}{\widehat{K}}^{\a\b\gamma\delta} 
  +\,\frac{3}{4}\, {^{(1)}_{\color{red}(1)}{\widehat {\Phi}}^{(\a\b}g^{\gamma\delta)}}
  + \frac{1}{8}\, {\phi}_{(1)}\,g^{(\a\b}g^{\gamma\delta)} \,,\\
  35&=&\hspace{16pt}25\hspace{15pt}\oplus\hspace{29pt}9\hspace{29pt}\oplus
\hspace{25pt}1\,.
\end{eqnarray}

Note, however, that the tensors (\ref{K2_SO3_1})-(\ref{K5_SO3_1}) are
still reducible.  In a next step for a finer decomposition, one has to
determine all tensors with symmetric or antisymmetric pairs or
triplets of indices. Furthermore, one has to consider tensors like
${\widehat{K}}^{\a[\b\g\d]}$ and their permutations, together with
their corresponding symmetric counterparts.  For example we find
$$ ^{(1)}{\widehat{K}}^{\a[\b\g\d]}=\,
^{(2)}{\widehat{K}}^{\a[\b\g\d]} 
=\,^{(3)}{\widehat{K}}^{\a[\b\g\d]}=\,
^{(4)}{\widehat{K}}^{\a[\b\g\d]}=0\,, $$
$$ \quad {\rm and}\quad
^{(5)}{\widehat{K}}^{\a[\b\g\d]}\neq 0\,, \, ^{(6)}{K}^{\a[\b\g\d]}\neq 0\,.
$$
In an accompanying paper we will work out a complete list of those
tensors.

\subsection{Decomposition of $K^{\a\b\g\d}[C]$ in Riemann--Cartan
  space}

The Kummer tensor density ${\cal K}^{\a\b\g\d}[{\cal T}]$ has 136, see
(\ref{irr-canonical}), and ${\cal K}^{\a\b\g\d}[{}^{(1)}{\cal T}]$
only 76 independent components, see (\ref{spec21}). Since the
principal part ${}^{(1)}{\cal T}^{\a\b\g\d}$ has 20, but the Weyl
tensor $C^{\a\b\g\d}$ merely 10 independent components, see the
decomposition (\ref{Rdecomp}), $K^{\a\b\g\d}[C]$ must then have
appreciably fewer independent components than 76. If we find its
irreducible decomposition, we should be able to determine them.

In analogy to Table 1, we list in Table 3 the
contractions of the Kummer-Weyl tensor
$K^{\a\b\gamma\delta}[C]$.\bigskip

\centerline{\bf Table 3. Contractions of {\em
    K\hspace{-1pt}ummer--W\hspace{-1pt}eyl}}\vspace{-10pt}
\begin{equation*}\label{Tab_CCC}
\begin{tabular}{|c||c|}\hline
  & $K^{\a\b\gamma\delta}[C] =  C^{\mu\a\lambda\b}\,
  ^{\star}C^{\rstar}{}_{\hspace{-3pt}\mu\rho\lambda\sigma}C^{\rho\gamma\sigma\delta}$
  \cr\hline\hline
  $^{(1)}K^{\a\b\gamma\delta}$ & $^{(1)}K^{\mu}{}_{\mu}{}^{\a\b} = \,
  ^{(1)}_{{\color{red}(1)}}{\Phi}^{\a\b}= \cdots =\,
  ^{(1)}_{{\color{red}(6)}}{\Phi}^{\a\b}=: K^{\a\b}$
  \cr\hline
  & $ K:= \, ^{(1)}_{{\color{red}(1)}}{\Phi}^{\mu}{}_{\mu} = \, ^{(1)}K^{\mu}{}_{\mu}{}_{\lambda}{}^{\lambda} $ \cr\hline\hline
  $^{(2)}K^{\a\b\gamma\delta}$ & $^{(2)}K^{\a\b\gamma\delta}=0$\cr\hline\hline
  $^{(3)}K^{\a\b\gamma\delta}$ & $^{(3)}_{{\color{red}(1)}}{\Phi}^{\a\b}=\,
  ^{(3)}_{{\color{red}(2)}}{\Phi}^{\a\b}= \, ^{(3)}_{{\color{red}(4)}}{\Phi}^{\a\b}= \,
  ^{(3)}_{{\color{red}(5)}}{\Phi}^{\a\b}$\,,
  \cr\hline
  & $^{(3)}_{{\color{red}(3)}}{\Phi}^{\a\b}=\,
  ^{(3)}_{{\color{red}(6)}}{\Phi}^{\a\b}= -
  \,^{(3)}_{{\color{red}(1)}}{\Phi}^{\a\b}/2$
  \cr\hline\hline
  $^{(4)}K^{\a\b\gamma\delta}$ & $^{(4)}_{{\color{red}(2)}}{\Phi}^{\a\b}=\,
  ^{(4)}_{{\color{red}(1)}}{\Phi}^{\a\b}\,,$
  \cr\hline
  & $^{(4)}_{{\color{red}(4)}}{\Phi}^{\a\b}=\, ^{(4)}_{{\color{red}(5)}}{\Phi}^{\a\b}
  = - \, ^{(4)}_{{\color{red}(1)}}{\Phi}^{\a\b}\,,$ \cr\hline
  & $^{(4)}_{{\color{red}(3)}}{\Phi}^{\a\b}=\,
  ^{(4)}_{{\color{red}(6)}}{\Phi}^{\a\b}= 0$ \cr\hline\hline
  $^{(5)}K^{\a\b\gamma\delta}$ & $^{(5)}K^{\a\b\gamma\delta}=0$
  \cr\hline\hline
  $^{(6)}K^{\a\b\gamma\delta}$ & $K^{[\a\b\gamma\delta]}\neq 0$
  \cr\hline\hline
\end{tabular}
\end{equation*}
We find only one independent first contraction $K^{\a\b}$ of the
Kummer-Weyl tensor, and, in turn, only one independent scalar $K$ as
second contraction. We choose $\{ K^{\a\b}\,, K \}$ as independent
objects for both contractions.

Moreover, we can additionally relate irreducible pieces to each other:
\begin{eqnarray}
  -2\, ^{(3)}_{{\color{red}(1)}}{\Phi}^{\a\b}  =   ^{(1)}_{{\color{red}(1)}}{\Phi}^{\a\b} \,, 
  \qquad -2\, ^{(4)}_{{\color{red}(1)}}{\Phi}^{\a\b}  = 
  ^{(1)}_{{\color{red}(1)}}{\Phi}^{\a\b}\,.
\end{eqnarray}
Note that all contractions are symmetric, that is, we do not have antisymmetric
pieces of ${\Phi}^{\a\b}$,
\begin{equation}
^{(I)}_{{\color{red}(A)}}{\Phi}^{[\a\b]}=0\, \quad
(I=1,\cdots ,5\quad A=1,\cdots ,6)\,.
\end{equation}
Table 3 allows also to prove interrelations between contractions of
the various irreducible pieces as, for example,
\begin{eqnarray}\label{Ksymfr}
  _{{\color{red}(1)}}{\widehat K}^{\a\b}[C] & := &
  K^{\mu}{}_{\mu}{}^{\a\b}-\frac{1}{4}\,  K^{\mu}{}_{\mu}{}^{\lambda}{}_{\lambda}
  g^{\a\b}
  =  \,^{(1)}_{{\color{red}(1)}}{\Phi}^{\a\b} + \, ^{(1)}_{{\color{red}(3)}}{\Phi}^{\a\b}
  + \, ^{(1)}_{{\color{red}(4)}}{\Phi}^{\a\b}\cr && -\frac{1}{4}\left[
    ^{(1)}_{{\color{red}(1)}}{\Phi}^{\mu}{}_{\mu} + 
\, ^{(1)}_{{\color{red}(3)}}{\Phi}^{\mu}{}_{\mu} + \,
    ^{(1)}_{{\color{red}(4)}}{\Phi}^{\mu}{}_{\mu}  \right]g^{\a\b}  =  0\,.
\end{eqnarray}
The five remaining contractions, defined analogously, fulfill the
trace-free condition, that is, for the full Kummer--Weyl tensor we have
\begin{equation}
{_{{\color{red}(A)}}{\widehat K}^{\a\b}[C] = 0\, \quad (A=1,\cdots ,6)}\,.
\end{equation}


\section{Outlook}

In order to win more insight into the properties of the Kummer tensor,
we will apply it to exact vacuum solutions of general relativity (GR)
and of the Poincar\'e gauge theory of gravitation (PG). In particular,
we will investigate the Kerr solution of GR and the Kerr metric {\it
  with torsion} \cite{MC_Bae_Gue_NC_1987,McKerr:1988} within PG. It
could be that the considerations of Burinskii \cite{Burinskii:2012jf},
who finds Kummer surfaces inside a Kerr black hole, have some relation
to our investigations.

In current string theory, the Kummer surface plays a role as a special
case of a so-called K3 (Kummer-K\"ahler-Kodaira) surface, see
\cite{Griffith:1978}. In turn, a K3 surface is a special case of a
3-dimensional complex Calabi-Yau manifold, see
\cite{Polchinski2:1998}. These manifolds are used for the
compactification of higher dimensions. However, in this application,
light propagation does not seem to play a role.

Quite generally, we wonder whether one can also attach a Kummer-like
tensor of rank four to the K3 surfaces and the Calabi-Yau manifolds
in differential geometry and in string theory, respectively.

\section{Acknowledgments} 

FWH is very grateful to {\it H.~Kn\"orrer} (Z\"urich) for advice on
Kummer surfaces and on algebraic geometry before this project came
under way.  FWH and YI acknowledge support of the German-Israeli
Research Foundation (GIF) by the grant GIF/No.1078-107.14/2009.  AF is
grateful for a fellowship of the German Academic Exchange Service
(DAAD) for a one-year stay in Cologne. He would like to thank the
Institute for Theoretical Physics, University of Cologne, for
hospitality where much of his contribution has been done. Moreover, AF
is grateful for an invitation to the University of Bremen, where he had
interesting and extended discussions with {\it Claus L\"ammerzahl} and
{\it Volker Perlick.}  AF also thanks {\it Matias Dahl} (Aalto) for
comments. We all are grateful to the referee for a number of very
useful suggestions.

\section{References}\vspace{-40pt}



\begin{thebibliography}{99}
\bibitem{Post} E.J.~Post, {\it Formal Structure of Electromagnetics}
  -- General Covariance and Electromagnetics. North Holland, Amsterdam
  (1962) and Dover, Mineola, NY (1997).

\bibitem{Birkbook} F.~W.~Hehl and Yu.~N.~Obukhov, {\it Foundations of
    Classical Electrodynamics: Charge, flux, and metric.}
  Birkh\"auser, Boston (2003).

\bibitem{Hehl:2007ut} F.~W.~Hehl, Yu.~N.~Obukhov, J.~-P.~Rivera and
  H.~Schmid, Relativistic nature of a magnetoelectric modulus of
  Cr$_2$O$_3$ crystals: A four-dimensional pseudoscalar and its
  measurement, Phys.\ Rev.\ A {\bf 77} (2008) 022106 
%
[\href{http://arXiv.org/pdf/0707.4407}{arXiv:0707.4407}].

\bibitem{Serdyukov:2001} A.~Serdyukov, I.~Semchenko, S.~Tretyakov, and
  A.~Sihvola, {\em Electromagnetics of Bi-anisotropic Materials,}
  Theory and Applications, Gordon and Breach, Amsterdam (2001).

 \bibitem{Ismobook} I.~V.~Lindell, {\it Differential Forms in
     Electromagnetics.} IEEE Press, Piscataway, NJ, and
   Wiley-Interscience (2004).

\bibitem{Sihvola-meta} A.~Sihvola, Metamaterials in electromagnetics,
  Metamaterials {\bf 1} (2007) 2--11.

\bibitem{DarrigolOptics} O.~Darrigol, {\it A History of Optics,} from
  Greek antiquity to the nineteenth century. Oxford University Press,
  Oxford (2012).

\bibitem{Knoerrer} H.~Kn\"orrer, {\it Die Fresnelsche Wellenfl\"ache.}
  In \cite{Knoerrer*} (1986) pp.115--141.

\bibitem{Hadamard} J.~Hadamard, {\it Lessons on the Propagation of
    Waves and the Equations of Hydrodynamics,} [translated from the
  French original ``Le\c{c}ons sur la propagation des ondes et les
  \'equations de l'hydrodynamique'' Hermann, Paris (1903)]
  Birkh\"auser, Basel (2011).

\bibitem{Schaefer} C.~Schaefer, {\it Einf\"uhrung in die theoretische
    Physik,} Vol.3, Part 1, de Gruyter, Berlin (1932).

\bibitem{Lindell1973} I.~V.~Lindell, Coordinate independent dyadic
    formulation of wave normal and ray surfaces of general anisotropic
    media, J.\ Math.\ Phys.\ {\bf 14} (1973) 65--67.

\bibitem{KiehnFresnel} R.~M.~Kiehn, G.~P.~Kiehn, and J.~B.~Roberds,
  Parity and time-reversal symmetry breaking, singular solutions, and
  Fresnel surfaces, Phys.\ Rev.\ A {\bf 43} (1991) 5665--5671.

\bibitem{Itin:2010} Y.~Itin, Dispersion relation for electromagnetic
   waves in anisotropic media,  {Phys.\ Lett.\ A} {\bf 374} (2010)
   1113--1116.

\bibitem{O'Dell} T.H.~O'Dell, {\it The Electrodynamics of
    Magneto-Electric Media.} North-Holland, Amsterdam (1970).

\bibitem{Bateman_1910} H.~Bateman, {Kummer's quartic surface as a wave
    surface}, Proc.\ London Math.\ Soc.\ Ser.2, {\bf 8}(1) (1910)
  375--382.

 \bibitem{Whittaker} E.\ Whittaker, {\it A History of the Theories of
    Aether and Electricity}, 2 volumes, reprinted. Humanities Press,
  New York (1973).

\bibitem{Sihvola} A.~H.~Sihvola, Are non-reciprocal bi-isotropic media
  forbidden? IEEE, Transactions on Microwave Theory and Techniques
  {\bf 43} (1995) 2160--2162.

\bibitem{RaabSihvola} R.~E.~Raab and A.~H.~Sihvola, On the existence
  of linear non-reciprocal bi-isotropic (NRBI) media, J.\ Phys.\ A
  {\bf 30} (1997) 1335--1344.

\bibitem{Lindell-Sihvola} I.~V.~Lindell and A.~H.~Sihvola, Perfect
   electromagnetic conductor, J.\ of Electromagnetic Waves and
   Applications {\bf 19} (2005) 861--869.

\bibitem{Obukhov:2004}
  Yu.~N.~Obukhov and F.~W.~Hehl,
  On possible skewon effects on light propagation,
  Phys.\ Rev.\ D {\bf 70} (2004) 125015
  [\href{http://arXiv.org/pdf/physics/0409155}{arXiv:physics/0409155}].

\bibitem{Itin:2013} Y.~Itin, Skewon no-go theorem, Phys.\ Rev.\ D
   {\bf 88} (2013) 107502.

\bibitem{Schouten_1954} J.~A.~Schouten, {\it Ricci-Calculus,} 2nd
  ed. Springer, Berlin (1954).

\bibitem{Itin:2009aa} Y.~Itin, {On light propagation in premetric
    electrodynamics. Covariant dispersion relation,} J.\ Phys.\ A {\bf
    42} (2009) 475402
  [\href{http://arXiv.org/pdf/0903.5520}{arXiv:0903.5520}].

\bibitem{Obukhov/Itin} Yu.~N.~Obukhov, Obukhov's proof of the
  equivalence of the dispersion relations, see Appendix of Itin
  \cite{Itin:2009aa}  (2009) pp.16--19.

\bibitem{Schuller:2009hn} F.~P.~Schuller, C.~Witte and
  M.~N.~R.~Wohlfarth, {Causal structure and algebraic classification
    of area metric spacetimes in four dimensions,} Ann.\ Phys.\ (NY)
  {\bf 325} (2010) 1853--1883
  [\href{http://arXiv.org/pdf/0908.1016}{arXiv:0908.1016}].

\bibitem{Lindell:2005} I.~V.~Lindell, {Electromagnetic wave equation
     in differential-form representation,} Progress in
   Electromagnetics Research (PIER) {\bf 54} (2005) 321--333.

\bibitem{Favaro:2012} A.~Favaro, {\it Recent Advances in Classical
 Electromagnetic Theory.} Ph.D. thesis, Imperial College London (2012).

\bibitem{Kummer:1864} E.~E.~Kummer, {\"Uber die Fl\"achen vierten
    Grades mit sechzehn singul\"aren Punkten [On surfaces of fourth
    order with sixteen singular points]} (1864). In:
  \cite{Weil}, pp.418--432.

\bibitem{Kummer:1864a} E.~E.~Kummer, {\"Uber die Strahlensysteme,
    deren Brennfl\"achen Fl\"achen vierten Grades mit sechzehn
    singul\"aren Punkten sind [On the ray systems, the focal surfaces
    of which are surfaces of fourth degree with sixteen singular
    points]} (1864). In: \cite{Weil}, pp.433--439.

\bibitem{LordProjGeom} E.~Lord, {\it Symmetry and Pattern in
     Projective Geometry.} Spinger, London (2013).

\bibitem{Kummer:1863} E.~E.~Kummer, {\"Uber atmosph\"arische
    Strahlenbrechung [On atmospheric refraction of rays]} (1863). In:
  \cite{Weil}, pp.337--349.

 \bibitem{Hudson_1903} R.~W.~H.~T.~Hudson, {\it Kummer's quartic
    surface.}
  Cambridge University Press, Cambridge (1903).

\bibitem{Cayley} A.~Cayley, In: {\it The Collected Mathematical Papers
    of Arthur Cayley}, Vol.I. Cambridge University Press, Cambridge (1889).
  Sur la surface des ondes, pp.302--305 [originally
  published in 1846].

\bibitem{Mathematica} S.~Wolfram, {\it MATHEMATICA,} Wolfram Research,
  Inc., {\it Mathematica Edition: Version 9.0}, Champaign, IL (2013).

\bibitem{Rocchini} C.~Rocchini (Florence), Kummer surface (2013). See
  the following link to Wikipedia:
  [\href{http://en.wikipedia.org/wiki/File:Kummer_surface.png}
  {Rocchini $(2013)$}].  \vspace{-5pt}

http://en.wikipedia.org/wiki/File:Kummer\underline{\hspace{6pt}}surface.png.

\bibitem{Rowe} D.~E.~Rowe, {Mathematical models as artefacts for
    research: Felix Klein and the case of Kummer surfaces,}
  Math. Semesterber. (Springer) {\bf 60} (2013) 1--24.

\bibitem{Delphenich:premetric}
  D.~H.~Delphenich, Symmetries and pre-metric electromagnetism,
  Annalen der Physik (Berlin) {\bf 14} (2005) 663--704
[\href{http://arXiv.org/pdf/gr-qc/0508035}{arXiv:gr-qc/0508035}].

\bibitem{Delphenich:projectiveI}
  D.~H.~Delphenich, Line geometry and electromagnetism I: basic
  structures, 26 pages,
%
[\href{http://arXiv.org/pdf/1309.2933}{arXiv:1309.2933}].

\bibitem{Delphenich:projectiveII}
  D.~H.~Delphenich, Line geometry and electromagnetism II: wave
  motion, 31 pages,
%
  [\href{http://arXiv.org/pdf/1311.6766}{arXiv:1311.6766}].

\bibitem{Jessop} C.~M.~Jessop, {\it A Treatise on the Line Complex.}
  Cambridge University Press, Cambridge (1903)
  [\href{http://archive.org/details/treatiseonlineco00jessuoft}
  {Jessop $(1903)$}].

\bibitem{ruse1944sets} H.~S.~Ruse, {Sets of vectors in a V$_4$ defined
    by the Riemann tensor,} J.\ London Math.\ Soc.\ {\bf 19} (1944)
  no.75, part 3, 168--178.

\bibitem{Ruse_1944b} H.~S.~Ruse, {On the line-geometry of the Riemann
    tensor}, Proc. Royal Soc. Edinburgh A {\bf 62} (1944) 64--73.

\bibitem{Zwillinger} D.~Zwillinger, S.~G.~Krantz, and K.~H.~Rosen
  (eds.), CRC Standard Mathematical Tables and Formulae,
  $31^{\hspace{-1pt}\text{st}}$ edition. Chapman \& Hall/CRC, Boca
  Raton (2003).

\bibitem{Avritzer:2007} D.~Avritzer and H.~Lange, Moduli spaces of
  quadratic complexes and their singular surfaces, Geometriae Dedicata {\bf
    127} (2007) 177--197.

 \bibitem{FavaroHehl:2012} A.~Favaro and F.~W.~Hehl, Fresnel versus
 Kummer surfaces: geometrical optics in dispersionless linear
 (meta)materials and vacuum. Invited lecture at {\it Electromagnetic
 Spacetimes,} Wolfgang Pauli Institute, Vienna, 19--23 Nov 2012,
[\href{http://arXiv.org/pdf/1401.4077}{arXiv:1401.4077}].

\bibitem{PRs} F.~W.~Hehl, J.~D.~McCrea, E.~W.~Mielke, and Y.~Ne'eman,
  {Metric-affine gauge theory of gravity: Field equations, Noether
    identities, world spinors, and breaking of dilation invariance},
  {Phys. Repts.} {\bf 258} (1995) 1--171.

\bibitem{Reader} M.~Blagojevi\'c and F.~W.~Hehl (eds.), {\it Gauge
    Theories of Gravitation, a Reader with Commentaries.} Imperial
  College Press, London (2013).

\bibitem{Obukhov:2006gea} Yu.~N.~Obukhov, Poincar\'e gauge gravity:
  selected topics, {Int.\ J.\ Geom.\ Meth.\ Mod.\ Phys.} {\bf 3}
  (2006) 95--138
  [\href{http://arXiv.org/pdf/gr-qc/0601090}{arXiv:gr-qc/0601090}].

\bibitem{Penrose} R.~Penrose and W.~Rindler, {\it Spinors and
    Space-Time,} Vols.I and II. Cambridge University Press, Cambridge
  (1984/86).

\bibitem{S_K_2004} H.~Stephani, D.~Kramer, M.~A.~H.~MacCallum,
  C.~Hoenselaers, and E.~Herlt, {\it Exact Solutions to Einstein's
    Field Equations,} 2nd ed. Cambridge University Press, Cambridge
  (2004).

\bibitem{LL} L.~D.~Landau and E.~M.~Lifshitz: {\it The Classical
    Theory of Fields}, Vol.2 of {\it Course of Theoretical Physics},
  p.\ 281; transl. from the Russian. Pergamon, Oxford (1962).

\bibitem{Zund_1969} J.~D.~Zund, {Algebraic iuvariants and the
    projective geometry of spinors,} Ann. di Matem. Pura et Appl. {\bf
    82} (1969) 381--412.

\bibitem{Ruse_1936} H.~S.~Ruse, {On the geometry of the
    electromagnetic field in general relativity}, Proc.\ London Math.\
  Soc.\ Ser.2, {\bf 41}(1) (1936)  302--322.

\bibitem{Hehl:2005hu} F.~W.~Hehl and Yu.~N.~Obukhov, {Spacetime
      metric {}from local and linear electrodynamics: A new axiomatic
      scheme,} Lect.\ Notes Phys.\ (Springer) {\bf 702} (2006)
    163--187 
    [\href{http://arXiv.org/pdf/gr-qc/0508024}{arXiv:gr-qc/0508024}].

\bibitem{Dahl:2011he} 
  M.~F.~Dahl,
  Non-dissipative electromagnetic medium with a double light cone,
  Annals of Physics (NY) {\bf 330} (2013) 55--73
 [\href{http://arXiv.org/pdf/1108.4207}{arXiv:1108.4207}].

\bibitem{DahlFavaro2014} M.~F.~Dahl and A.~Favaro, Extracting the
  electromagnetic response of local and linear (meta)materials from
  the optical phase velocity: the case of two light cones. Forthcoming
  (2014/15).

\bibitem{Gilkey} P.~B.~Gilkey, {\it Geometric Properties of Natural
    Operators Defined by the Riemann Curvature Tensor.} World
  Scientific, River Edge, NJ (2001).

\bibitem{Zund_1976} J.~D.~Zund, {A m\'emoir on the projective geometry
    of spinors}, Ann. di Matem. Pura et Appl. {\bf 110} (1976)
  29--135.

\bibitem{Weyl} H.~Weyl, {\it The Classical Groups. Their Invariants
    and Representations,} 2nd ed.\ with supplements. Princeton
  University Press, Princeton, NJ (1953).

\bibitem{Boerner} H.~Boerner, {\it Representations of Groups.} North
  Holland, Amsterdam (1970).

\bibitem{hamermesh} M.~Hamermesh, {\it Group Theory and its
    Application to Physical Problems.} Dover, New York (1989).

\bibitem{wade} T.~L.~Wade, Tensor algebra and Young's symmetry
  operators, {American J.\ Math.} {\bf 63} (1941) 645--657.

\bibitem{Ma:2004} Z.-Q.~Ma and X.-Y.~Gu, {\it Problems and Solutions
    in Group Theory for Physicists.} World Scientific, Singapore
  (2004) p.204.

\bibitem{Johannes} J.~Grabmeier, E.~Kaltofen, and V.~Weispfenning
  (eds.), {\it Computer Algebra Handbook.} Foundations, Applications,
  Systems. Springer, Berlin (2003).

\bibitem{Hearn} A.~C.~Hearn, {\em {REDUCE} User's Manual, Version 3.5}
  RAND Publication CP78 (Rev. 10/93). The RAND Corporation, Santa
  Monica, CA 90407-2138, USA (1993). Nowadays Reduce is freely
  available for download; for details see
  [\href{http://reduce-algebra.com/}{reduce-algebra.com}] and
  [\href{http://reduce-algebra.sourceforge.net/}{sourceforge.net}].

\bibitem{Socorro} J.~Socorro, A.~Macias, and F.~W.~Hehl, {Computer
    algebra in gravity: Reduce-Excalc programs for (non-)Riemannian
    space-times.\ I,} {Comput.\ Phys.\ Commun.} {\bf 115} (1998)
  264--283
  [\href{http://arXiv.org/pdf/gr-qc/9804068}{arXiv:gr-qc/9804068}].

\bibitem{reduce93} D.~Stauffer, F.~W.~Hehl, N.~Ito, V.~Winkelmann,
  and J.\,G. Zabolitzky, \emph{Computer Simulation and Computer
    Algebra---Lectures for Beginners}, 3rd ed. Springer, Berlin
  (1993).

\bibitem{MC_Bae_Gue_NC_1987} J.~D.~McCrea, P.~Baekler, and M.~G\"urses,
  {A Kerr-like solution of the Poincar\'e gauge field equations},
  Nuovo Cimento B {\bf 99} (1987) 171--177.

\bibitem{McKerr:1988} P.~Baekler, M.~G\"urses, F.~W.~Hehl, and
  J.~D.~McCrea, {The exterior gravitational field of a charged
    spinning source in the Poincar\'e gauge theory: A Kerr-Newman
    metric with dynamic torsion,} {Phys.\ Lett.\ A} {\bf 128}
  (1988) 245--250.

\bibitem{Burinskii:2012jf} A.~Burinskii, {Complex structure of the
    four-dimensional Kerr geometry: Stringy system, Kerr theorem, and
    Calabi-Yau twofold,} Adv.\ High Energy Phys.\ {\bf 2013} (2013)
  509749
  [\href{http://arxiv.org/pdf/1211.6021.pdf}{arXiv:1211.6021}].

\bibitem{Griffith:1978} P.~Griffith and J.~Harris, {\it Principles of
    Algebraic Geometry.} Wiley, New York (1978).

\bibitem{Polchinski2:1998} J.~Polchinski, {\it String Theory, Volume
    II.} Cambridge University Press, Cambridge (1998).

\bibitem{Knoerrer*} H.~Kn\"orrer et al., {\it Mathematische
    Miniaturen, Vol. 3: Arithmetik und Geometrie,} vier Vorlesungen.
  Birkh\"auser, Basel (1986).

\bibitem{Weil} A.~Weil (ed.), {\it Ernst Eduard Kummer, Collected
    Papers,} Vol.II: Function theory, geometry and
  miscellaneous. Springer, Berlin (1975).

  %
  %
  %
%
%
%
%
%
%
%
%
%
%
%
%
\end{thebibliography}
\end{document}